\documentclass[preprint2]{aastex}

\shorttitle{Morphological correlations among B-fields, filament, and bipolar bubbles in RCW57A}
\shortauthors{Eswaraiah et al.}
\usepackage[maxfloats=40]{morefloats} 
\usepackage{color}
\usepackage{catoptions}
\usepackage[breaklinks,colorlinks,citecolor=blue,linkcolor=blue]{hyperref} 
\usepackage{hypcap}
\newcommand{\Rom}[1]{\uppercase\expandafter{\romannumeral #1\relax}}
\makeatletter
\usepackage{natbib,rotating,cite,multicol}
\usepackage{pdflscape}
\usepackage{lscape}
\usepackage{amssymb}
\usepackage{longtable}
\usepackage{endnotes}
\usepackage{helvet}
\usepackage{setspace}
\usepackage{amsmath}
\usepackage{breqn}

\def\hii{H {\sc ii}~}

\def\rmn{\rm}

\def\kms{km\,s$^{-1}$}

\begin{document}

\title{Understanding the links among magnetic fields, filament, the bipolar bubble, and star formation in RCW57A using NIR polarimetry}
\author{Chakali Eswaraiah\altaffilmark{1,2,3}, Shih-Ping Lai\altaffilmark{1}, Wen-Ping Chen\altaffilmark{2}, A. K. Pandey\altaffilmark{3}, M. Tamura\altaffilmark{4,5}, G. Maheswar\altaffilmark{3}, S. Sharma\altaffilmark{3}, Jia-Wei Wang\altaffilmark{1}, S. Nishiyama\altaffilmark{6}, Y. Nakajima\altaffilmark{7}, Jungmi Kwon\altaffilmark{8}, R. Purcell\altaffilmark{9}, and A. M. Magalh{\~a}es\altaffilmark{10}}
\altaffiltext{1}{Institute of Astronomy, National Tsing Hua University, 101 Section 2, Kuang Fu Road, Hsinchu 30013, Taiwan}
\altaffiltext{2}{Institute of Astronomy, National Central University, 300 Jhongda Rd, Jhongli, Taoyuan Country 32054, Taiwan}
\altaffiltext{3}{Aryabhatta Research Institute of Observational Sciences (ARIES), Manora-peak, Nainital, Uttarakhand-state, 263002, India}
\altaffiltext{4}{Department of Astronomy, Graduate School of Science, The University of Tokyo, 7-3-1 Hongo, Bunkyo-ku, Tokyo 113-0033, Japan}
\altaffiltext{5}{National Astronomical Observatory of Japan, 2-21-1 Osawa, Mitaka, Tokyo 181-8588, Japan}
\altaffiltext{6}{Miyagi University of Education, 149 Aramaki--aza--Aoba, Aoba--ku, Sendai, Miyagi 980-0845, Japan}
\altaffiltext{7}{Center of Information and Communication Technology, Hitotsubashi University, 2-1 Naka, Kunitachi, Tokyo 186-8601, Japan}
\altaffiltext{8}{Institute of Space and Astronautical Science, Japan Aerospace Exploration Agency, 3-1-1 Yohinodai, 
Chuo-ku, Sagamihara, Kanagawa 252-5210, Japan}
\altaffiltext{9}{Sydney Institute for Astronomy (SIfA), School of Physics, The University of Sydney, NSW 2006, Australia}
\altaffiltext{10}{Inst. de Astronomia, Geofisica \& Ciencias Atmosfericas, Univ. de Sao Paulo, Sao Paulo, Brazil}
\email{eswarbramha@gmail.com}

\begin{abstract}
The influence of magnetic fields (B-fields) in the formation and evolution of bipolar bubbles, due to the expanding 
ionization fronts (I-fronts) driven by the H{\sc ii} regions that are formed and embedded 
in filamentary molecular clouds, has not been well-studied yet. 
In addition to the anisotropic expansion of I-fronts into a filament, 
B-fields are expected to introduce an additional anisotropic pressure which 
might favor expansion and propagation of I-fronts to form a bipolar bubble. 
We present results based on near-infrared polarimetric observations 
towards the central $\sim8\arcmin\times8\arcmin$ area of the star forming region RCW57A which hosts an 
H{\sc ii} region, a filament, and a bipolar bubble. 
Polarization measurements of 178 reddened background stars, 
out of the 919 detected sources in the $JHK_{s}$-bands, reveal 
B-fields that thread perpendicular to the filament long axis. The B-fields exhibit an hour-glass morphology 
that closely follows the structure of the bipolar bubble. 
The mean B-field strength, estimated using the Chandrasekhar-Fermi method, is 91$\pm$8~$\mu$G. 
B-field pressure dominates over turbulent and thermal pressures. Thermal pressure might act in the same orientation as those of B-fields 
to accelerate the expansion of those I-fronts. The observed morphological correspondence among 
the B-fields, filament, and bipolar bubble 
demonstrate that the B-fields are important 
to the cloud contraction that formed the filament, gravitational collapse and star formation in it, 
and in feedback processes. The latter include the formation and evolution of mid-infrared bubbles by 
means of B-field supported propagation and expansion of I-fronts. 
These may shed light on preexisting conditions favoring the formation of the massive 
stellar cluster in RCW\,57A. 
\end{abstract}

\keywords{Polarization - (ISM:) dust, extinction - ISM: magnetic fields - open clusters and associations: individual: RCW57A}

\section{\sc Introduction}\label{intro}

Massive stars ($>$~8M$_{\sun}$) have profound effects on their surrounding natal cloud material through 
driving strong ionizing radiation, powerful stellar winds, outflows, expanding H{\sc ii} regions, 
and supernova explosions which either trigger, or halt, 
star formation \citep{Beutheretal2002,ZinneckerYorke2007,LeeChen2007,Smithetal2010,Roccatagliataetal2013}. 
Star forming regions associated with H{\sc ii} regions exhibit
spectacular morphologies, such as spherical, or ring-like bubbles, and 
bipolar or unipolar bubbles \citep{Churchwelletal2006,Churchwelletal2007,Deharvengetal2010,Simpsonetal2012} that may also 
be surrounded by bright rimmed clouds (BRCs) \citep{Sugitanietal1991,Sugitanietal1994}. 
Studies based on data from recent space missions, such as {\it Herschel}, {\it Spitzer}, and {\it WISE}, 
have shown that most molecular clouds exhibit filamentary structure 
\citep[c.f., ][]{Andreetal2010,Arzoumanianetal2011,Arzoumanianetal2013,Hilletal2011,Perettoetal2012,Benedettinietal2015} 
and, furthermore, that they are associated with bipolar bubble-like features\footnote{Bipolar bubbles are also referred to as hourglass-shaped nebulae, bipolar nebulae, bi-lobed appearance nebulae, a bipolar H{\sc ii}
regions \citep[e.g., ][]{StaudeElsasser1993,Minieretal2013,Deharvengetal2015}} \citep[e.g., ][]{Minieretal2013,Deharvengetal2015,ZhangCPetal2016}.
Examples of molecular cloud filaments associated with
bipolar/unipolar bubbles include N131 \citep{ZhangCPetal2016},
Sh2-106 \citep{Ballyetal1998}, IRS16/Vela\,D \citep{Strafellaetal2010},
NGC\,2024 \citep{Mezgeretal1992}, Sh2-201 \citep{Mampasoetal1987}, Sh2-88 \citep{WhiteFridlund1992}, 
RCW\,36 \citep{Minieretal2013}, and RCW\,57A \citep{Minieretal2013}. 

Though bipolar bubbles are the natural outcome of anisotropic 
expansion of ionizing fronts from H{\sc ii} regions hosting 
massive O/B-type star(s) located in filaments \citep{Minieretal2013,Deharvengetal2015,ZhangCPetal2016}, 
details involved in their formation and evolution processes remain poorly understood. 
According to 2D \citep{Bodenheimeretal1979} and 3D \citep{FukudaHanawa2000} 
hydrodynamic simulations, the evolution of an H{\sc ii} region in a filamentary cloud induces supersonic and subsonic I-fronts 
along the minor and major axes of the filament, respectively. This results the distribution of low and high density 
material along the minor and major axes. Thus, the I-fronts experience more hindered flows 
along the major axis than along the minor axis. As a consequence of this anisotropic expansion, 
a bipolar bubble will be formed in a filament (Figure 1 of \citealt{Deharvengetal2015}). 
Though a few studies \citep[e.g.,][]{Minieretal2013,Deharvengetal2015} were devoted to 
the study of bipolar bubbles, there exists no study focused on exploring the connection between the 
B-fields anchored in the filaments and their role in the formation 
and evolution of bipolar bubbles. 
\citet{FukudaHanawa2000} did include B-fields 
(parallel to the filament long axis) in their simulations, 
but they studied only the influence of B-fields on shape, separation, and formation epoch of the cores 
that formed along the filament.

B-fields are believed to guide the contraction of cloud material to form filaments in molecular clouds,
and thus play a crucial role in controlling cloud stability and collapse, fragmentation into cores, 
and internal star formation 
\citep{PereyraMagalhaes2004,Alvesetal2008,Lietal2013,PlanckCollaboration2014,FrancoAlves2015}.
A few studies (e.g., \citealt{PereyraMagalhaes2007} towards the IRAS Vela Shell and \citealt{Wisniewskietal2007} 
towards supershell NGC\,2100~in the LMC) 
suggest the importance of B-fields in the dynamical evolution of 
the expanding shells driven by O/B-type stars. 
The interplay between the expansion of an ionized nebula and the influence of B-fields 
has been the subject of several studies \citep[e.g.,][]{PavelClemens2012,Santosetal2012,Santosetal2014,PlanckcollXXXIVAlvesetal2015,Kwonetal2010,Kwonetal2011,Kwonetal2015}. 
These studies hint that B-fields permeated through filamentary molecular clouds 
would also influence the expanding I-fronts or 
outflowing gas, from an H{\sc ii} region created in the filament. 

\begin{figure*}[!ht]
\centering
	\resizebox{12.0cm}{13cm}{\includegraphics{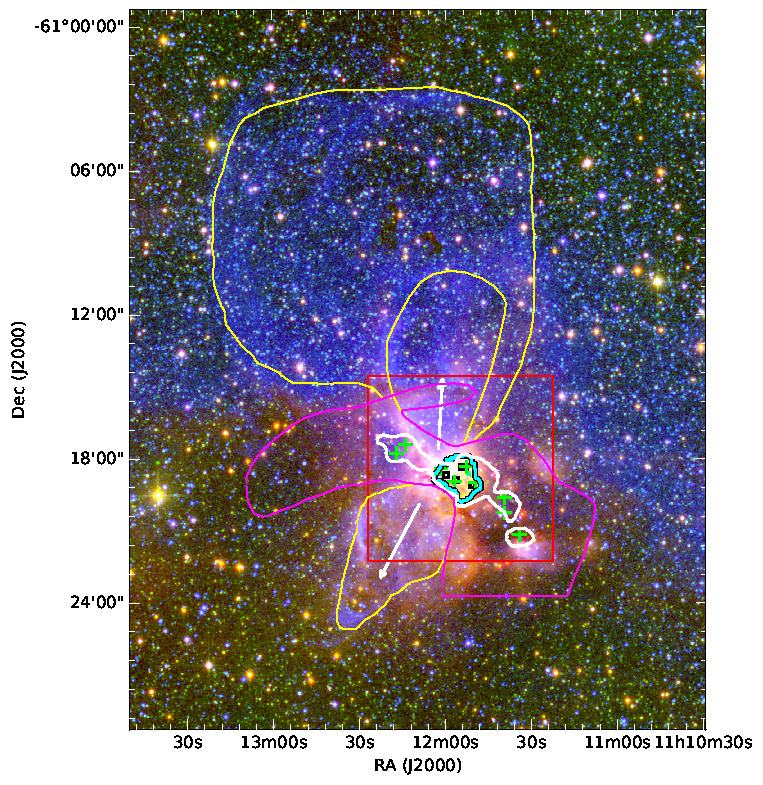}}
\caption{Tricolor image of the RCW57A region made using {\it WISE} 4.6~$\mu$m (red), 2MASS $K_{s}$-band (green), and
DSS2 $R$-band (blue) images. Various contours, extracted from \citet{Purcelletal2009}, are overplotted. 
Yellow contours enclosing white arrows represent the bipolar 
bubble-like structure observed in mid-infrared ({\it MSX}, {\it Spitzer}, and {\it WISE}) images. This 
feature appears as an extended widened loop in the $R$-band, as shown with an extended 
yellow contour in the Northern part. Thick cyan contour represents the 3.4\,cm radio free-free emission from ionized gas, 
delimiting the extent of the H{\sc ii} region. The white contours 
represent the 1.1\,mm dust continuum emission \citep{Hilletal2005}, showing an elongated dusty filament
from NE to SW which crosses the \hii region. 
The extent of the molecular cloud, from $^{13}$CO(1~--~0) line emission \citep{Purcelletal2009}, is 
depicted with magenta outline (the horizontal line at the bottom of the magenta outline is the limit of 
the mapped area). Infrared sources \citep[IRS;][]{FrogelPersson1974} 
and water/methanol maser sources \citep[cf.][]{Purcelletal2009} are shown with open black squares and green crosses, respectively. 
Possible directions of the outflowing gas \citep[cf.][]{dePreeetal1999} are shown with white arrows.
Red square box denotes the area of $7\farcm7\times7\farcm7$ observed with SIRPOL. 
}
\label{WISE4p6_K_R_color_composite}
\end{figure*}

RCW57A (also known as NGC\,3576, G291.27-0.70, or IRAS\,11097-6102) is an H{\sc ii} 
region associated with a filament and bipolar bubble and is 
located at a distance of 2.4~--~2.8 kpc \citep{Persietal1994,dePreeetal1999}. 
We adopt 2.4 kpc, which is within uncertainties of both the kinematic and spectroscopic determinations 
\citep[see][]{Persietal1994}. Figure \ref{WISE4p6_K_R_color_composite} 
depicts the overall morphology of RCW57A.
It contains optically bright 
nebulosity with several dark globules and luminous arcs \citep{Persietal1994}. 
It is one of the massive star forming regions in the southern sky, 
hosting an H{\sc ii} region (cyan contour) embedded in a filament (white contour) from which a 
widely-extended bipolar bubble (yellow contours) 
is emerging. A deeply embedded near-IR cluster, consisting of more than 130 young stellar objects (YSOs) is
associated with this region \citep{Persietal1994}. 
The observed ratios of the infrared fine-structure ionic lines 
(Ne {\sc ii}, Ar {\sc iii}, and S {\sc iv}) \citep{Lacyetal1982} 
indicate that at least eight O7.5V stars are necessary to account for the ionization of the region. 
However, even these stars may not be sufficient to account for the Ly$\alpha$ ionizing photons 
inferred from radio data \citep{Figueredoetal2002,Barbosaetal2003,Townsley2009}. 
Based on the newly discovered cluster of stars, using X-ray data, \citet{Townsleyetal2014} 
suggested that an additional cluster of OB stars might be deeply embedded that were not known before because of 
heavy obscuration. This cluster is located slightly NW of the center of the near-IR cluster. 
The 10~$\mu$m map \citep[cf., ][]{FrogelPersson1974}, reveals 
the presence of five infrared sources (IRS; black squares in Figure \ref{WISE4p6_K_R_color_composite}) 
near the center of the H{\sc ii} region. These, together with water and methanol 
maser sources (green crosses in Figure \ref{WISE4p6_K_R_color_composite}) distributed along the filament, are 
indicative of active ongoing star formation in RCW57A \citep{FrogelPersson1974,Caswell2004,Purcelletal2009}. 
Therefore, RCW\,57A is an ideal target to investigate the morphological links among 
filaments, bipolar bubbles, and B-fields so as to understand the star formation history. 

Here, near-infrared (NIR) polarimetric observations towards the central region of RCW57A have been carried out 
to map the plane-of-sky B-field geometry in the star forming region. 
When unpolarized background starlight passes through an interstellar cloud, 
dust particles partially aligned by a 
B-field linearly polarize the light by a few percent. Virtually all possible mechanisms 
responsible for the dust grain alignment with respect to the B-fields 
\citep{DavisGreenstein1951,Anderssonetal2015,Lazarianetal2015book} 
yield directions of the polarized light tracing the average B-field orientation projected on the plane-of-sky. 

The aim of this paper is to understand whether (a) the B-field structure in the molecular cloud guides 
feedback processes, such as expansion and propagation of outflowing gas and ionization fronts, which lead
to the formation of bipolar bubbles, or (b) feedback processes regulate the resulting B-field structure.
The B-field is {\it active} in the former scenario, while it is {\it passive} in the latter.

The structure of this paper is as follows. 
Section \ref{data_reduction} describes the observations and data reduction. 
Detailed analyses are presented in Section \ref{yso_removal}, the main purpose of which 
is to identify and remove the YSOs from the list of foreground and background stars by utilizing 
NIR and mid-infrared (MIR) colors, NIR polarization measurements, 
and polarization efficiencies. Results are presented in Section \ref{sec_results}. 
This section presents estimates of the B-field strength, and detailed 
comparisons between B-field pressure to turbulent and thermal pressures. 
A possible evolutionary scenario, based on the morphological 
correspondences among the filament, bipolar bubble, and B-fields is 
discussed in Section \ref{sec_discuss}. 
Furthermore, based on the scenario that best characterizes our observed B-fields in RCW57A, 
we discuss the possible preexisting conditions that might favored the
formation of star cluster in RCW57A. 
Conclusions are summarized in Section \ref{conclusions}.

\section{\sc Observations and data reduction}\label{data_reduction}

Simultaneous observations in $J$- (1.25~$\mu$m), $H$- (1.63~$\mu$m), and $K_{s}$- (2.14~$\mu$m) bands towards the 
central star forming region of RCW57A ($\alpha$~$=~11^{\rmn h}11^{\rmn m}54\fs8$, $\delta$~$=$~$-61\arcdeg18\arcmin26\arcsec$ [J2000]) 
were carried out on 2007 May 6 using the imaging polarimeter SIRPOL 
(polarimetric mode of the Simultaneous IR Imager for Unbiased Survey (SIRIUS) camera: \citealt{Kandorietal2006}), 
mounted on the IR Survey Facility (IRSF) 1.4-m telescope at the South Africa Astronomical Observatory (SAAO). 
The SIRIUS camera was equipped with a rotating achromatic (1~--~2.5~$\mu$m) half-wave plate (HWP) 
and high extinction-ratio wire grid analyzer, three 1024$\times$1024 HgCdTe (HAWAII) 
IR detectors and $JHK_{s}$ filters for simultaneous observations \citep{Nagshimaetal1999,Nagayamaetal2003}. 
The field of view was $7\farcm7\times7\farcm7$ with a pixel scale of 0$\farcs$45 pixel$^{-1}$.

One set of observations consisted of 10 s exposures at four HWP 
position angles (0$\degr$, 22.5$\degr$, 45$\degr$ and 67.5$\degr$) 
at 10 dithered sky pointings. Such sets of 4$\times$10~images were repeated towards the same sky coordinates to increase 
signal-to-noise. Sky frames were also obtained between target observations. 
The total integration time was 400 s per HWP angle. The average seeing was $1\farcs54$ ($J$), 
$1\farcs44$ ($H$) and $1\farcs31$ ($K_{s}$). 

Master flats were created by utilizing the evening and twilight flat-field frames on the same night of the observations. 
We processed the data using the dedicated data reduction pipeline
`{\sc pyIRSF}'\footnote{{\sc pyIRSF} package uses {\sc PyRAF}, {\sc python}, and {\sc c} language scripts. 
This pipeline was used, specifically, for the reduction of NIR polarimetric data acquired
with SIRPOL on IRSF and was written and compiled by Yasushi Nakajima. 
Detailed description of the pipeline software and their individual tasks can be
found at \url{https://sourceforge.net/projects/irsfsoftware/}. {\sc PyRAF} is a product of the 
Space Telescope Science Institute, which is operated by AURA for NASA.}. 
This pipeline used the raw data from the target, sky, dark, and flat field frames as inputs. 
The data reduction tasks included dark subtraction, flat-field correction, median sky subtraction, frame registration, and 
averaging. The final products of this pipeline were the average combined four 
intensity images ($I_{0}$, $I_{22.5}$, $I_{45}$, and $I_{67.5}$) 
corresponding to the four positions of the HWP. 

\subsection{\it Aperture photometry of point sources}\label{Secaperphot}

We performed aperture photometry of point-like sources on the intensity images for the 
HWP angles ($I_{0}$, $I_{22.5}$, $I_{45}$, and $I_{67.5}$) of each of the $JHK_{s}$-bands using 
{\sc iraf}\footnote{{\sc iraf} is distributed by the US National Optical Astronomy Observatory,
which is operated by the Association of Universities for Research in Astronomy, Inc., under 
a cooperative agreement with the National Science Foundation.} and the {\sc idl} Astronomy Library \citep{Landsman1993}. 
Point sources with peak intensities above 5$\sigma$ (where $\sigma$ is the rms uncertainty 
of the particular image pixel values) of the local sky background were detected 
using {\sc DAOFIND}. Aperture photometry was performed using the 
{\sc PHOT} task of the {\sc DAOPHOT} package in {\sc IRAF}. 
An aperture radius was chosen to be nearly the FWHM of the point-like sources, i.e., 3.4, 3.2, and 2.9 pixels 
in the $J$-, $H$-, and $K_{s}$ bands, respectively. The sky annulus inner radius was set to 
be 10 pixels with a 5 pixel width. 
A source that matched in center-to-center positions to within 1 pixel radius 
(i.e., 2 pixel diameter equivalent to $\sim1\arcsec$) 
in all four HWP images was judged to be same star. 

The Stokes $I$ intensity of each point-like source was calculated using 
\begin{equation}\label{eqnstokeI}
I= \frac{1}{2}(I_{0}+I_{22.5}+ I_{45}+I_{67.5})
\end{equation}
and used to estimate its photometric magnitude. There were 
401 stars found with both IRSF and 2MASS $JHK_{s}$ photometric magnitude uncertainties less than 0.1 mag. 
These stars were used to calibrate the IRSF instrumental magnitudes and colors to the 2MASS system. 
Details regarding this calibration can be found in Appendix \ref{appendix_a}. 
From the HWP images, we were able to estimate $JHK_{s}$ photometric magnitudes and colors ($[J-H]$ and $[H-K_{s}]$)
for 702 stars to uncertainties of less than 0.1 mag.

\subsection{\it Aperture polarimetry}\label{aperpol}

Polarimetry of point-like sources (aperture polarimetry) 
was performed on the combined intensity images at each HWP angle. 
Extracted intensities were used to estimate the Stokes parameters of each star using 
\begin{eqnarray}\label{stokes_params_eqns}
Q &=& (I_{0}-I_{45}), \\
U &=& (I_{22.5}-I_{67.5}). 
\end{eqnarray}
The aperture and sky radii were the same as those used in the aperture photometry on total intensity ($I$) images.
To obtain the Stokes parameters ($Q$ and $U$) in the equatorial coordinate system, 
an offset rotation of 105$\degr$ \citep{Kandorietal2006,Kusuneetal2015} was applied.
We calculated the degree of polarization $P$ and the 
polarization angle $\theta$ as follows:
\begin{eqnarray}\label{stoke_eq1}
P&=&\frac{\sqrt{Q^2+U^2}}{I} 
\end{eqnarray}
\begin{eqnarray}\label{stoke_eq2}
\theta&=&\frac{1}{2} \arctan\left(\frac{U}{Q}\right)
\end{eqnarray}
The errors in $P$ and $\theta$ were also estimated by propagating the errors in Stokes parameters according to equations \ref{stoke_eq1} and \ref{stoke_eq2}. 
Since the polarization degree, $P$, is a positive definite quantity, the derived
$P$ value tends to be overestimated, especially for low $S/N$ cases. To correct for this bias, we calculated
the de-biased $P$ using $P_{db}$=$\sqrt{P^2-\delta P^2}$ \citep{WardleKronberg1974}, where $\delta P$ is the error in $P$.
It should be noted here that, hereafter, we consider $P_{db}$ as $P$. 
The absolute accuracy of the offset polarization angle for SIRPOL was estimated to be
less than 3$\degr$ \citep{Kandorietal2006}. The polarization efficiencies of the wire grid polarizer 
are 95.5\%, 96.3\%, and 98.5\% at the $J$-, $H$-, and $K_{s}$- bands, respectively, 
and the instrumental polarization is less
than 0.3\% over the field of view at each band \citep{Kandorietal2006}. Due to these high polarization efficiencies
and low instrumental polarization, no additional corrections were made to the data. 
To obtain the astrometric plate solutions,
we matched the pixel coordinates of the detected sources and the celestial coordinates of their counterparts in the 2MASS 
Catalog \citep{Skrutskie2006}, and applied {\it ccmap}, {\it ccsetwcs} and {\it cctran} tasks of the {\sc iraf}
{\it imcoords} package to the matched lists. The rms error in the returned coordinate system was $\sim$0.03$\arcsec$.
However, since the 2MASS astrometry system is limited about to 0.2$\arcsec$, the stellar 
positional uncertainties should take that into account. A total of 919 stars had polarization detection in at least one band with $P/\sigma_{P} \geq$~1.

\subsection{\it Complete data sample}\label{comp_data_sample}
 
The polarimetric data for the 919 stars were merged with the photometry of the 702 stars, and 
the resultant data are presented in Table \ref{polphotdata1}. 
Star IDs are given in column 1. 
The photometric data of the stars, those do not have them from IRSF, are extracted 
from 2MASS \citep{Skrutskie2006} and are denoted with an asterisk along with star IDs in column 1. 
{The coordinates, aperture polarimetric and photometric results of each source 
are presented in columns 2~--~12.}
Column 13 reports our classifications of the stars, as described in section 
\ref{yso_removal} below. Though we have presented all the stellar data 
satisfying the $P/\sigma_{P}\,\geq\,1$ criterion in Table \ref{polphotdata1}, 
further analyses used only the 461 stars having $H$-band polarization data 
which satisfied $P(H)/\sigma_{P(H)}~\geq~2$ criterion. 

\begin{figure}[!ht]
\centering
\resizebox{7.5cm}{16cm}{\includegraphics{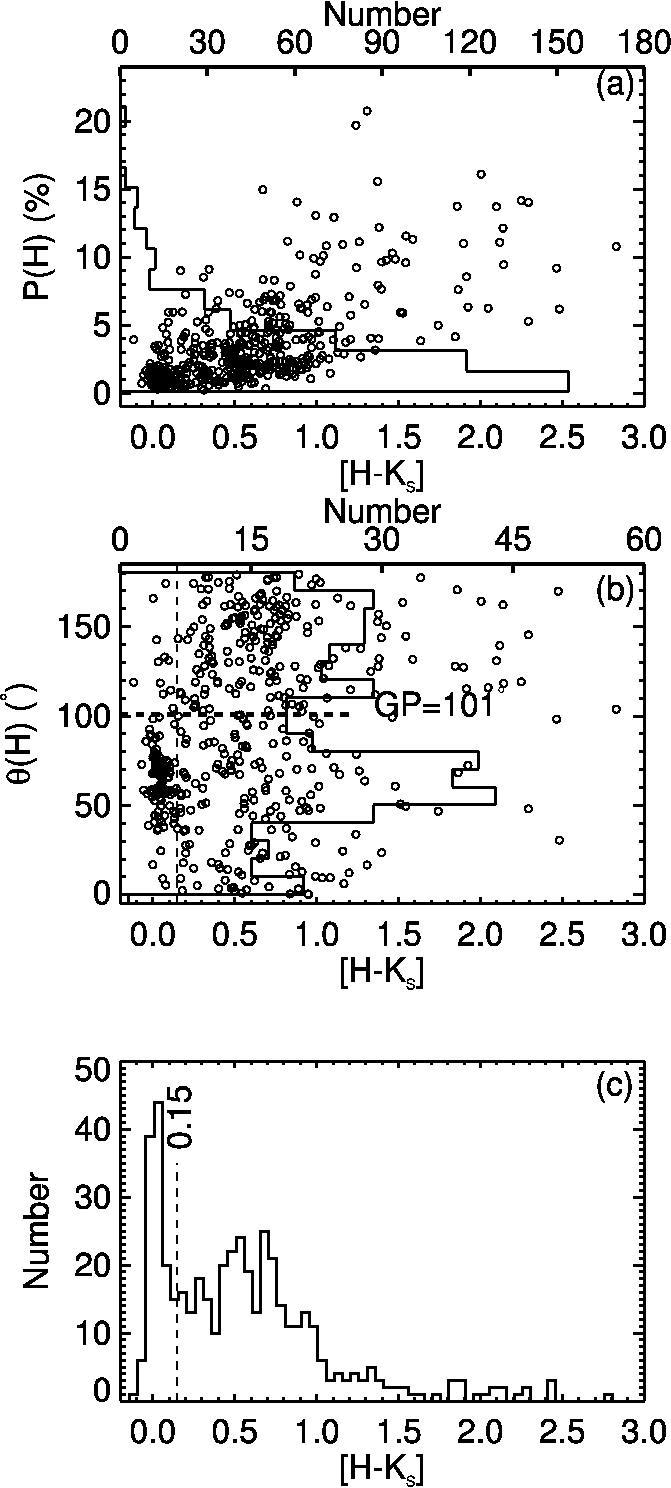}}
\caption{(a) $P(H)$ and (b) $\theta(H)$ versus $[H-K_{s}]$ color for 461 stars. 
The histograms of $P(H)$ (a), $\theta(H)$ (b), and $[H-K_{s}]$ (c) are also shown.} 
\label{p_pa_vs_hk}
\end{figure}

Distributions of $P(H)$ versus $[H-K_{s}]$ and $\theta(H)$ versus $[H-K_{s}]$ for the 461 stars are, respectively, presented in Figures \ref{p_pa_vs_hk}(a) \& (b). 
Also shown are the histograms of $P(H)$ (Figure \ref{p_pa_vs_hk}a), $\theta(H)$ (Figure \ref{p_pa_vs_hk}b), 
and $[H-K_{s}]$ (Figure \ref{p_pa_vs_hk}c). 
$[H-K_{s}]$ colors plotted in Figure \ref{p_pa_vs_hk} 
(also in Figures \ref{ppv_nirccd}, \ref{pol_effi_j_h_k}, and \ref{nirccd_and_ppahk_bffg}) 
were converted to the California Institute of Technology (CIT)
system using the relations given by \citet{Carpenter2001}. The degree of polarization $P(H)$ exhibits an 
increasing trend with $[H-K_{s}]$ color. 
The histogram of $P(H)$ peaks at $\sim$~1\% of polarization. 
The polarization angles $\theta(H)$ of the stars full span~0~--~180$\degr$ range with two groups densely concentrated 
around $\theta(H)$~$\simeq$~60$\degr$ and $\theta(H)$~$\simeq$~150$\degr$. These two dominant $\theta(H)$ distributions seem to be different 
from the position angle of 101$\degr$ corresponding to the orientation of the 
Galactic plane (GP) at $b$~$=$~$-$0.72$\degr$, 
shown with a dashed line. 
Moreover, these two group of stars exhibit different distributions of $[H-K_{s}]$ colors 
and are separated by a boundary 
at $[H-K_{s}]$~$=$~0.15 mag. While these two groups of stars exhibit different properties of 
$\theta(H)$ and $[H-K_{s}]$ colors, they 
are not as clearly separated from each other in the distribution of $P(H)$. 

\begin{figure}[!ht]
\centering
\resizebox{7.25cm}{7.5cm}{\includegraphics{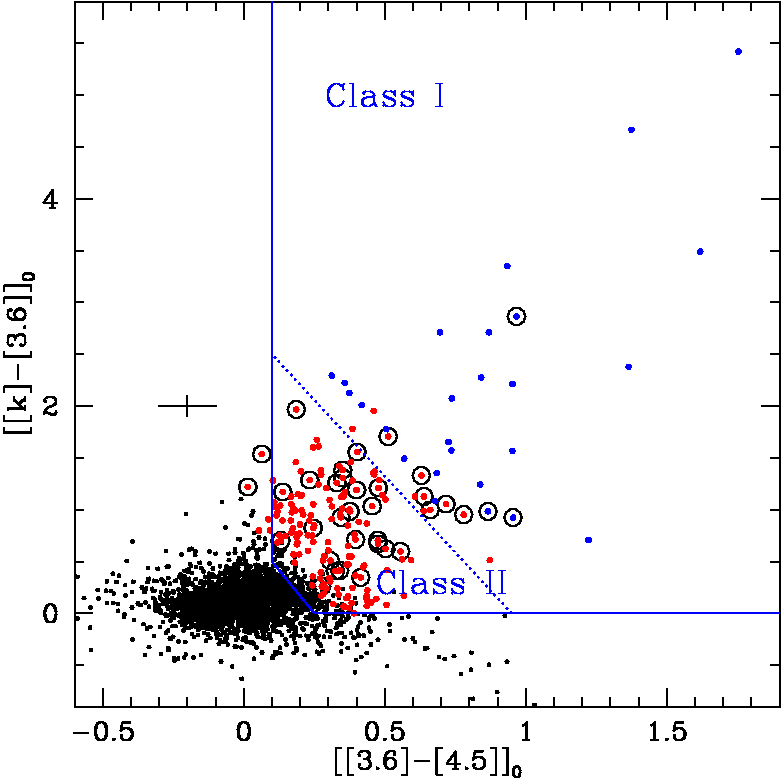}}
	\caption{$[[K_{s}]-[3.6]]_{0}$ vs. $[[3.6]-[4.5]]_{0}$ TCD (two color diagram) for the {\it Spitzer} IRAC sources.
Black dots are the total sample in the entire region. 
The blue and red filled circles represent Class I and Class II sources, respectively in the RCW57A region. 
The black circles surrounding some of the points are three Class I and thirty two Class II sources 
having $2\sigma$ polarization measurements. 
The slanted dotted line denotes the boundary between Class I and Class II YSOs. The average error in the colors is shown.}
\label{spitzer_ccd}
\end{figure}

\section{\sc Identification of YSOs and foreground stars}\label{yso_removal}

The B-field inside the molecular cloud can be probed by using polarimetry of 
background stars whose light becomes differentially extincted and linearly polarized by the dust in 
the cloud. Hence, a sample of background stars needs to be identified among all the stars with measured polarizations. 
For this purpose, we identify and exclude the following sources: 
YSOs (Class I, Class II, HAe/Be stars and NIR excess sources) based on MIR and NIR color-color diagrams, 
stars with $L$-band excess using the data from \citet{Maerckeretal2006}, 
stars projected against the nebulous region where their polarizations might be 
contaminated by scattering (especially at the cluster center), 
the stars with non-interstellar polarization origin due to reflected light 
from their circumstellar disks/envelopes (whose 
properties are not discerned based on MIR and NIR colors), and the stars 
that have suffered depolarization effects due to multiple B-field components 
(associated with multiple dust layers) along their lines of sight. 
The latter two are identified using polarization efficiency diagram. 

Below, we present MIR and NIR two color diagrams (TCDs) and polarization efficiency ($P$ versus $H-K_{s}$) diagram 
to identify and exclude the above mentioned stars with intrinsic polarization. 
Finally, the remaining sample, that are free from intrinsic polarization, 
is sub-categorized further into foreground (FG) and background (BG) stars, 
based on their NIR colors and polarization characteristics. 

\subsection{\it YSOs identified by their mid-infrared colors}\label{mir_tcd}

Since the RCW57A region is still enshrouded in its natal molecular cloud, as evident from the $^{13}$CO(1~--~0) map 
(magenta contour in Figure \ref{WISE4p6_K_R_color_composite} and red background in Figure \ref{polvecmap_bg}), YSOs in the region can be 
deeply embedded. Therefore, the {\it Spitzer} MIR observations were used to probe deeper insight into the embedded YSOs. These 
occupy distinct regions in the {\it Spitzer} IRAC color plane, which makes MIR TCDs a useful tool for the classification of YSOs. Since
8.0~$\mu$m data are not available for the region, we used $[[K_{s}]-[3.6]]_{0}$ and $[[3.6]-[4.5]]_{0}$ \citep[cf.][]{Gutermuthetal2009} 
to identify deeply embedded YSOs, as shown in Figure \ref{spitzer_ccd}. 
The zones of Class I and Class II YSOs, based on the color criteria of \citet{Gutermuthetal2009}, are also depicted. 
Three sources were identified as Class I and thirty two sources as Class II and so were excluded from analyses. 
Identified Class I and Class II sources are mentioned in column 13 of Table \ref{polphotdata1}.

\subsection{\it YSOs identified by their near-infrared colors}\label{nir_tcd}

\begin{figure*}[!ht]
\centering
\resizebox{9.9cm}{12.5cm}{\includegraphics{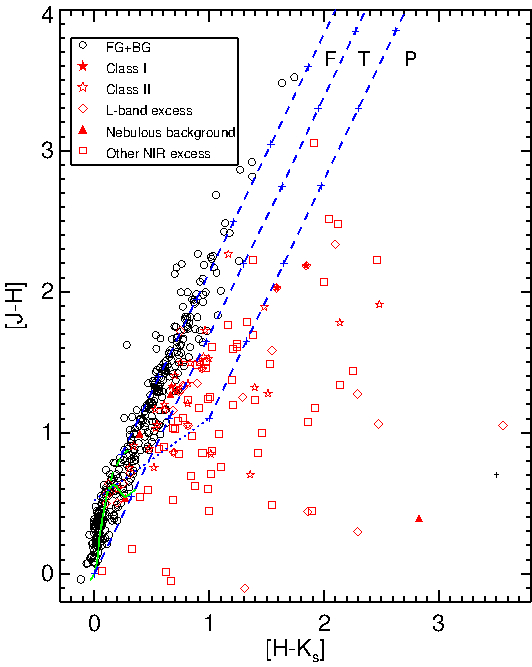}}
	\caption{
$[J-H]$ vs. $[H-K_{s}]$ NIR TCD (two color diagram) for the 461 stars sample. 
The green thin and thick continuous curves represent the unreddened MS and giant
branch \citep{BessellBrett1988}, respectively.
The dotted blue line indicates the locus of unreddened CTTSs. 
The parallel blue dashed lines are the reddening vectors drawn from the tip (spectral type M4)
of the giant branch (left reddening line), from the base (spectral type $A0$) of the MS branch
(middle reddening line), and from the tip of the intrinsic CTTS line (right reddening line) 
with the extinction ratios $A_{J}/A_{V}$~$=$~0.265, $A_{H}/A_{V}$~$=$~0.155, 
and $A_{K}/A_{V}$~$=$~0.090 \citep{Cohenetal1981}. The blue plus symbols on the 
reddening vectors show increments of $A_{V}$ by 5 mag each. 
The sources located in the ``F" region could be either reddened field stars 
or Class II and Class III sources with small NIR excesses. The sources distributed in the ``T" and ``P"
regions are considered to be mostly Classical T-Tauri stars \citep[CTTSs; ][]{MeyerCalvetHillenbrand1997}
or Class II sources with relatively large NIR excesses and likely Class I sources, respectively
\citep[for details see][]{Pandeyetal2008,Chauhanetal2011a,Chauhanetal2011b}. Typical 
errors in the plotted colors are comparable to the size of the symbols.} \label{ppv_nirccd}
\end{figure*}

NIR colors are also important tools to classify YSOs in star forming regions. 
Figure \ref{ppv_nirccd} shows the $[J-H]$ vs. $[H-K_{s}]$ NIR TCD for the 461 stars. 
The 461 stars were classified into several types as follow. 
The three Class I sources and 32 Class II sources 
identified based on the {\it Spitzer} MIR TCD (Figure \ref{spitzer_ccd}) are shown with filled and open asterisks, respectively. 
Fifty seven stars, falling in ``T" and ``P" regions and satisfying the relation $[J-H]$~$\textless$~1.69~$[H-K_{s}]$, are shown with squares 
exhibit NIR excesses. Four stars, denoted with filled triangles, are found to have nebulous backgrounds. 
Twenty eight stars, shown with open diamonds, were found to have $L$-band excesses \citep{Maerckeretal2006}.  
Two of the stars found to be Class I also have $L$-band excess, and similarly 3 stars found to be Class II also have $L$-band excess. 
Star identified as having NIR excess, $L$-band excess, and nebulous background are noted in column 13 of Table \ref{polphotdata1}.

The remaining 342 stars that appear to be free from intrinsic polarization and distributed in the zone ``F" in Figure \ref{ppv_nirccd} are 
(i) lightly reddened foreground field stars (FG), (ii) moderately reddened cluster members and highly reddened background stars (hereafter both cluster members and background stars referred to as BG), 
and (iii) weak line T-Tauri stars (WTTSs) or Class III sources. 
The locations of the WTTSs in the NIR TCD overlap with reddened background stars. 
Generally WTTSs exhibit almost negligible NIR-excess, as their disks will have been evaporated already or very 
optically thin \citep[eg., ][]{Adamsetal1987,AndrewsWilliams2005,Ciezaetal2007}. Therefore, 
it is reasonable to assume that these WTTSs are not intrinsically polarized by scattered light from the 
negligible amount of circumstellar material. 

In order to estimate the level of FG contamination, a
control field ($\alpha$~$=$~$11^{\rmn h}18^{\rmn m}12\fs22$, $\delta$~$=$~$-61\arcdeg59\arcmin38\farcs92$ [J2000]) of angular size 
$7\farcm7\times7\farcm7$ located $\sim$60$\arcmin$ from the RCW57A region was chosen.
Comparing the distributions of $[H-K_{s}]$ colors of the control field with colors in the RCW57A region 
shows that the  field stars to have $[H-K_{s}]$~$\textless$~0.15. 
Therefore, the 108 RCW57A stars with $[H-K_{s}]$ ~$\textless$~0.15 and lying in the `F' 
region of the NIR TCD are judged to be FG star candidates and remaining the 234 stars with $[H-K_{s}]$ ~$\geq$~0.15 and lying in the `F' region are 
judged to be BG star candidates. 

\begin{figure}[!ht]
\centering
\resizebox{7.5cm}{7.5cm}{\includegraphics{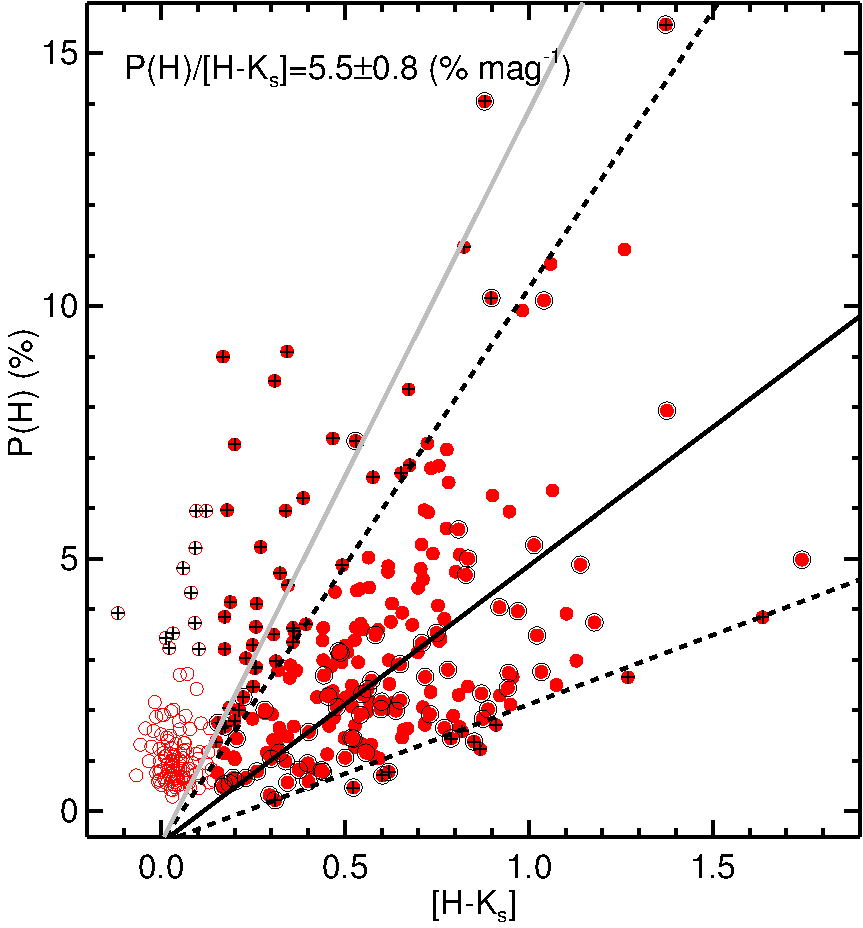}}
\caption{Polarization efficiency in $H$-band.
$P(H)$ versus $[H-K_{s}]$ for 108 FG candidate (open red circles) and 234 BG candidate (filled red circles) stars.
Encircled red filled circles correspond to the 76 BG candidate stars with 
$\sigma_{P(H)}$~$\textless$~0.3\%. 
Linear fit (solid lines) was made to these encircled stars, and the slope and uncertainty are quoted. 
Dashed lines are drawn with slopes twice and half of the slope
correspond to the solid lines. Plus symbols overlaid on open and filled circles are the stars 
excluded from the list of confirmed FG and BG stars.
Gray line in each panel represents the upper limit polarization
efficiency relation corresponds to $P(V)/E(B-V)$~$\leq$~9\%~mag$^{-1}$ \citep{Serkowskietal1975}.}. \label{pol_effi_j_h_k}
\end{figure}

\subsection{\it Final confirmed FG and BG stars based on polarization efficiency diagrams}\label{secpoleffi}

Polarization efficiency is defined as the ratio of polarization degree to interstellar reddening, such 
as $P(\lambda)/E(\lambda_{1}-\lambda_{2})$. 
Observed polarization efficiencies of dust grains are found to vary from one line of sight to the another. 
Reddening ($E(B-V)$) and polarization in $V$-band ($P_{V}$) are correlated and can be 
represented by the observational upper limit relation: $P_{V}/E(B-V)$~$\leq$~9\%~mag$^{-1}$ \citep{Serkowskietal1975}.
Polarization measurements that exceed this upper limit relation 
are generally attributed to the intrinsic polarization (non-magnetic polarization) 
caused by scattered light from dust grains in circumstellar disks or 
envelopes around YSOs \citep[e.g., ][]{Kandorietal2007,Eswaraiahetal2011,Eswaraiahetal2012,SeronNavarreteetal2016}. 

We also seek to identify and exclude the stars that suffer depolarization due 
to their starlight's passing through multiple B-field 
components along the line of sight. For optically thin clouds, $P(\lambda)$ is expected to be proportional to the 
column density if the field structure is uniform. 
In contrast, presence of multiple B-fields with different orientations along a line of sight will result depolarization \citep{Martin1974} and hence 
weakens the correlation between polarization and extinction \citep[eg.,][]{Eswaraiahetal2011,Eswaraiahetal2012}. 
Therefore, polarization efficiency diagrams 
\citep[eg.,][]{Kwonetal2011,Sugitanietal2011,Hatanoetal2013} will help identify 
stars with either intrinsic polarization or depolarization. 

Figure \ref{pol_effi_j_h_k} show $P$ versus $[H-K_{s}]$ diagrams of FG and BG candidate stars 
(Section \ref{nir_tcd}) with $H$-band data. 
These diagrams contain no known YSOs as we excluded them in the above Sections \ref{mir_tcd} and \ref{nir_tcd}. 
The BG candidate stars show an increasing trend in polarization in 
accordance with an increasing $[H-K_{s}]$ and 
well distributed below the upper limit interstellar polarization efficiency relation (gray line),
$P(H)/E(H-K_{s})$~$=$~14.5\%~mag$^{-1}$ corresponding to $P(V)/E(B-V)$~$=$~9\%~mag$^{-1}$ \citep[][also see \citealt{Hatanoetal2013} for 
detailed description on deriving polarization efficiency relations in NIR wavelengths.]{Serkowskietal1975}. 
To find the stars with non-magnetic polarizations (non interstellar origin of polarizations) and 
depolarization affect, first, we fit a straight line to polarization versus extinction for the BG stars 
with $\sigma_{P(H)}~\textless$~0.3\% (encircled filled circles). 
Second, we drew dashed lines with slope of twice (upper dashed line) 
and half (lower dashed line) of the fitted slopes. 
Third, we identified the stars falling above and below these lines 
as probable BG stars having influenced by non-magnetic origin of polarization 
and depolarization affect, respectively. The remaining stars 
falling between the two dashed lines are ascertained as confirmed BG stars. 
Therefore, the final number of confirmed BG stars 
was found to be 178~in $H$-band.  
Additional 64 candidate BG stars falling above and below the dashed lines (plus symbols) 
are mentioned as probable BG stars with intrinsic polarization or depolarization in Table \ref{polphotdata1}.

FG candidate stars (open circles) exhibit two distributions and those with 
$P(H)~\textgreater~3\%$ (plus symbols on open circles) may be affected by 
additional polarization caused by scattering or simply with larger uncertainties. 
We consider them as probable FG stars. Remaining 97 stars with $P(H)~\leq~3\%$ are 
considered as confirmed FG stars in $H$-band. 
These, confirmed and probable FG stars with $P(H)~\textgreater~3\%$ are mentioned in Table \ref{polphotdata1}. 

In the further analysis, we used only the confirmed 97 FG and 178 BG stars with $H$-band data.           
The fitted slope of $P(H)/E(H-K_{s})$\,$=$\,5.5$\pm$0.8\%\,mag$^{-1}$ (thick line) suggest that 
polarization efficiency of dust grains in the star forming cloud RCW57A are relatively higher than those of 
other Galactic line of sights \citep[e.g.,][]{Hatanoetal2013,Kusuneetal2015}. 

\begin{figure*}[!ht]
\centering
\resizebox{12cm}{12cm}{\includegraphics{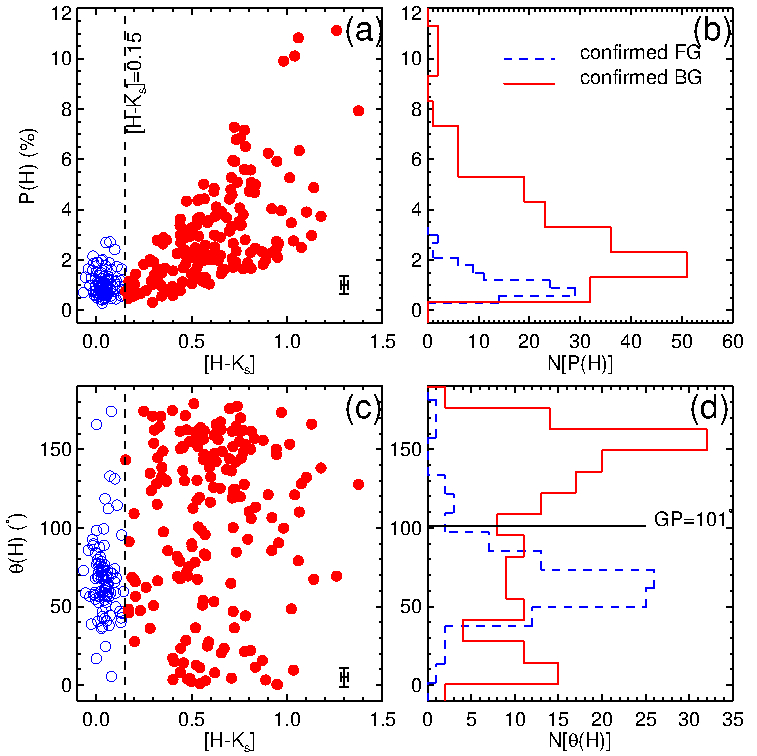}}
\caption{
Polarization properties of confirmed FG and BG stars. 
	(a) $P(H)$ versus $[H-K_{s}]$ for the 97 confirmed FG (open blue circles) and 178 
	confirmed BG (filled red circles). 
	(c) same as (a) but for $\theta(H)$ versus $[H-K_{s}]$. 
	Dashed line drawn at $[H-K_{s}]$~$=$~0.15, in (a) and (c), separates FG and BG stars. 
	Typical errors are plotted at the bottom right-corner of (a) and (c). 
	(b) Distributions $P(H)$ for the 97 confirmed FG (broken histogram) and 178 confirmed BG (continuous histogram) stars as plotted in (a) and (c). 
	(d) Same as (b) but for $\theta(H)$.} \label{nirccd_and_ppahk_bffg}
\end{figure*}

\section{\sc Results}\label{sec_results}

\subsection{\it Separation of foreground and background magnetic fields}\label{distri_p_t_of_fg_bg}
FG stars towards any distant cluster are, generally, 
less polarized and extincted than BG stars \citep{Eswaraiahetal2011,Eswaraiahetal2012,Pandeyetal2013}. 
This is because the dust layer(s) between the observer and the FG star contains fewer 
aligned dust grains, and because the degree of polarization 
and extinction are linearly correlated \citep{AannestadPurcell1973,Serkowskietal1975,Hatanoetal2013}. 
If the B-field orientation of the foreground 
dust layer is sufficiently different from that of the cloud, then the foreground stars would exhibit 
different polarizations ($P$ and $\theta$) from those of the background stars. 
Therefore, foreground and background stars can also be distinguished based on their polarization characteristics ($P$ vs $\theta$ or $Q$ versus $U$), 
as well as with their NIR colors \citep{Medhietal2008,Medhietal2010,Eswaraiahetal2011,Eswaraiahetal2012,Medhietal2013,Pandeyetal2013,Santosetal2014}. 
 
 Figures \ref{nirccd_and_ppahk_bffg}(a)--(d) depict that the 97 confirmed FG and 178 confirmed BG stars are well separated in their 
polarization characteristics and $[H-K_{s}]$ colors. 
Foreground stars have $[H-K_{s}]$~$\textless$~0.15, 
$P(H)$~$\textless$~3\% with a peak around 1.0$\%$, 
and a single Gaussian distribution of $\theta(H)$
(Gaussian mean of 65$\degr$ with a standard deviation of 14$\degr$). 
Background stars have $[H-K_{s}]~\geq$~0.15, 
$P(H)$ between 0.5~--~12\% with a peak $\sim 2\%$ and an
extended tail towards greater $P$ values and widely distributed $\theta(H)$ 
(range from 0$\degr$ to 180$\degr$) indicating the presence of complex B-field structure in RCW57A with 
a prominent peak at $\theta(H)$~$\sim$163$\degr$. About 55\% of stars are distributed between 110~--~180$\degr$,
whereas the remaining 45\% of stars are distributed between 0~--~110$\degr$.
This suggests that the plane-of-sky B-field structure of RCW\,57A 
is dominated by a component with $\theta(H)$~$\sim$163$\degr$, which is nearly orthogonal to the
foreground B-field component ($65\degr$) as well as being nearly orthogonal to the position
angle of the major axis of the filament ($\sim$60$\degr$).
Figure \ref{nirccd_and_ppahk_bffg}(d) also reveals that the distributions of $\theta(H)$ of
the confirmed FG and BG stars are different from the position angle (101$\degr$) of 
the Galactic plane (GP) at $b$~$=$~$-$0.72$\degr$, shown with a thick vertical line.

\subsection{\it B-field structure towards RCW57A}\label{vecmapsall}

\subsubsection{\it Contribution of foreground (FG) polarization}\label{vecmap_fg}

In order to account for the influence of FG polarization on the 
polarization angles of the confirmed BG stars, 
below we performed the FG subtraction.
First, the polarization measurements of confirmed 97 FG stars were converted into Stokes parameters. 
Second, weighted mean FG Stokes parameters were computed and found to be $Q_{FG}$~$=$~$-$0.332$\pm$0.008\% and 
$U_{FG}$~$=$~0.845$\pm$0.007\% (with standard deviation of 0.64\% in $Q_{FG}$ and 0.70\% in $U_{FG}$) 
and were subtracted vectorially from those 
of each confirmed BG star. Third, FG-subtracted Stokes parameters of the BG stars were then converted back to 
$\theta(H)_{{\rmn FG-corrected}}$.  
The Gaussian mean and standard deviation, for the distribution of
offset polarization angles of the BG stars ($\theta(H)~-~\theta(H)_{{\rmn FG-corrected}}$), are found to be 
1$\arcdeg$ and 7$\arcdeg$, respectively. 
Hence, there was no obvious change in the FG corrected 
B-field morphology inferred from the BG stars. 
Therefore, we ignored the FG contribution to the BG polarization measurements. 

\subsubsection{\it Magnetic field geometry}\label{vecmap_bg}
 
\begin{figure*}
\centering
\resizebox{14cm}{12cm}{\includegraphics{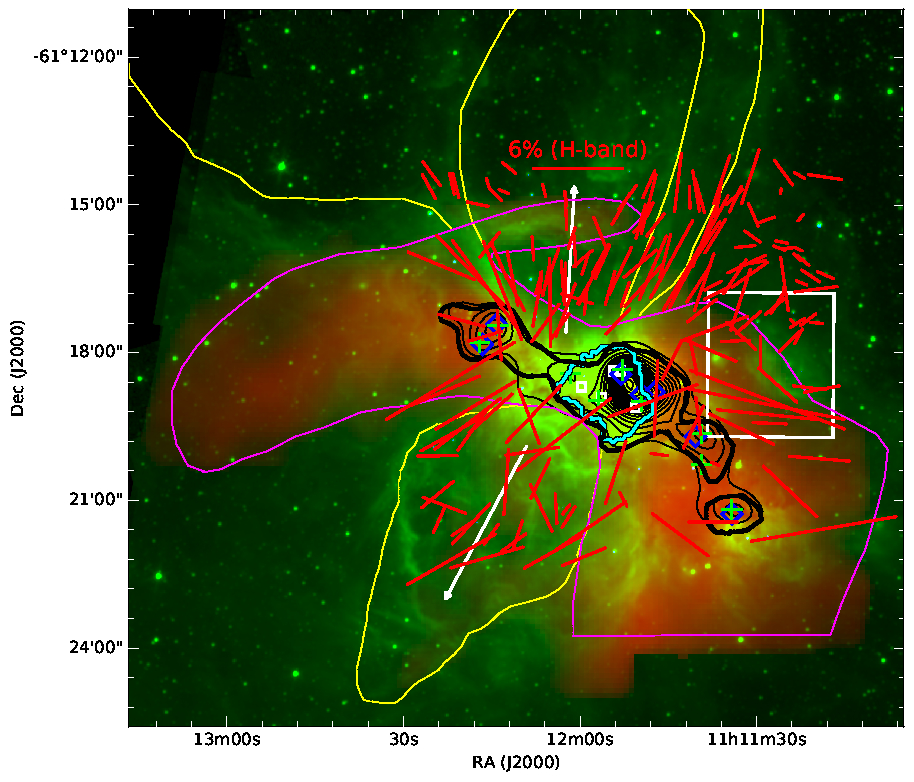}}
\caption{$H$-band polarizations (red vectors) of the 178 confirmed BG stars 
overlaid on the tricolor image constructed from the 
${^{13}}$CO(1~--~0) total integrated emission map (red) \citep{Purcelletal2009},
{\it Spitzer} IRAC Ch2+Ch3 combined image (green), and SIRPOL $J+H+K_{s}$-band combined image (blue). 
Reference vector with $P$~=~6$\%$ and $\theta$~=~90$\degr$ is plotted. 
Thick black contour represents SIMBA 1.2-mm dust emission \citep{Hilletal2005} with a flux level of $\sim$3 Jy/beam. 
Thin black contours denote ATLASGAL 0.87-mm thermal dust emission ranging from 2 Jy/beam to 22 Jy/beam 
(beam size of $18\farcs2$) with an interval of 1 Jy/beam. Cyan contour represent the extent of the H{\sc ii} region, 
as traced in 3.4-cm radio free-free emission \citep{dePreeetal1999}. 
Yellow contours delineate the morphology of the bipolar bubble. Magenta contour depicts the extent of the 
molecular cloud, as traced by $^{13}$CO(1~--~0) emission using {\it Mopra} telescope \citep{Purcelletal2009}. 
White square box shows the location of a Bright Rimmed Cloud. White arrows represent possible 
orientations of expanding I-fronts or 
outflowing gas. Green plus marks and black squares distributed 
along the filament are the water/methanol masers and IRS sources, respectively. Blue diamonds correspond to the 
seven massive cores (namely, S1-M1, S1-M2, S3-M4, S3-M5, S3-C3, S4-M6 and S5-M6) identified by \citet{Andreetal2008}. All the information on 
various contours, and the location of masers and IRS sources is extracted from \citet{Purcelletal2009}, by permission.\label{polvecmap_bg}}
\end{figure*}

Figure \ref{polvecmap_bg} shows the vector map of the $H$-band polarization measurements of the 
178 confirmed BG stars. 
The polarization vector map reveals a systematically ordered B-field that is 
configured into an hour-glass morphology which closely resembles 
the structure of the bipolar bubble. 

The direction of expanding I-fronts or outflowing gas is revealed based on a large NS velocity gradient in a radio 
recombination line observations \citep{dePreeetal1999}. These high velocity outflows signatures are
further traced in $^{13}$CO(1~--~0) position-velocity diagrams (Figure \ref{pvcuts}; shown below).
Shocked gas is visible at the SE and NW ends of the white arrows
in Figure \ref{polvecmap_bg} (also see Figure \ref{WISE4p6_K_R_color_composite}),
highlighting the outflow-cloud interaction. \citet[][see their Figures 3 and 4]{Townsleyetal2011a} have also showed that outflows from the 
deeply embedded protostars are responsible for the soft X-ray emission observed in the SE part of the cloud. 
These outflows have created the cavities depicted with enclosed
yellow contours (Figure \ref{WISE4p6_K_R_color_composite}).
Interestingly, the directions of the expanding I-fronts and/or outflowing gas, 
as depicted with white arrows (Figure \ref{polvecmap_bg}),
follow the B-field orientations.  

B-field are compressed along the rim of BRC (located inside of white square box in
Figure \ref{polvecmap_bg}) and more details of which are addressed in Appendix \ref{appendix_b}.

\subsection{\it Extracting essential parameters for estimating magnetic field strength}\label{esti_b}

To test whether B-fields play an active role in the formation of bipolar bubbles, 
it is essential to estimate the B-field strength by using the \citet{ChandrasekharFermi1953} relation 
(hereafter CF Method). For this purpose, below we determine the 
dispersion in $\theta(H)$ for the background stars, the gas velocity dispersion using CO data, and 
the dust and gas volume density using {\it Herschel} data.

\subsubsection{\it Dispersion in polarization angles: $\sigma_{\theta(H)}$}

\begin{figure}[!ht]
\centering
        \resizebox{7.5cm}{6.0cm}{\includegraphics{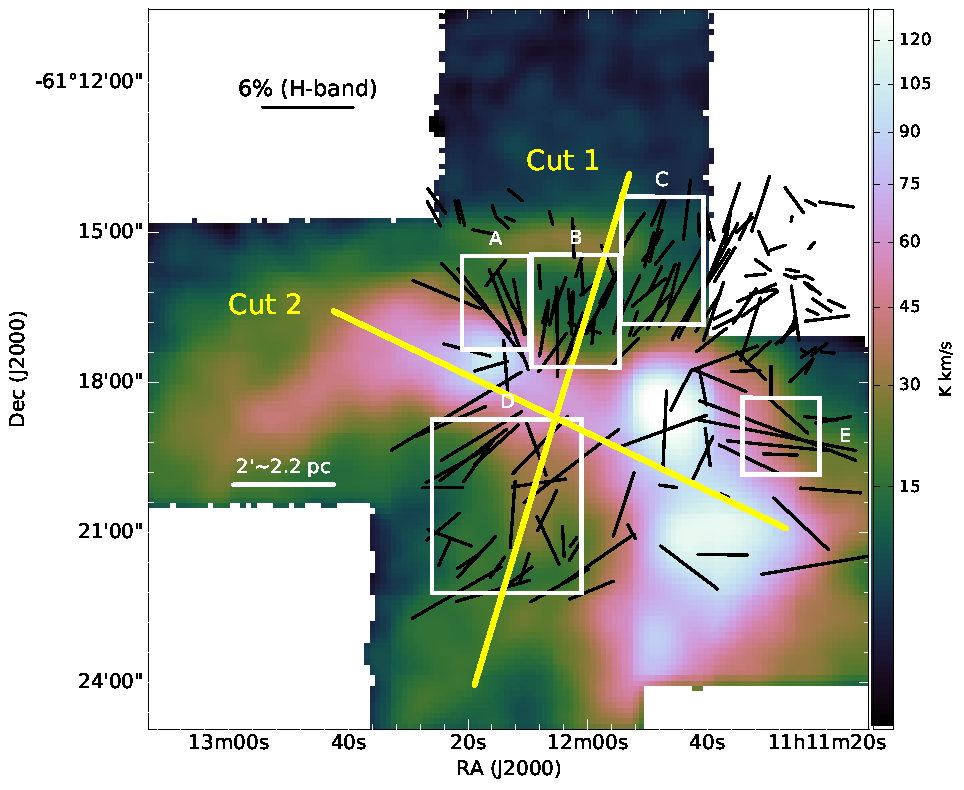}}
	\caption{$H$-band polarization vectors (black lines) of confirmed BG stars are overlaid on the 
	$^{13}$CO(1~--~0) \citep{Purcelletal2009} total integrated intensity image. 
	The units of the color scale are K~\kms. White rectangles, identified as A, B, C, D and E, 
	are the five regions selected for estimating the dispersions in gas velocity and $\theta(H)$. 
	Yellow thick lines represent `Cut 1' and `Cut 2' along the ionization front or outflowing gas 
	and cloud main axis, respectively, as described in the text and used as the basis for Figure \ref{pvcuts}. 
	A reference length scale with 2$\arcmin$ corresponding to 2.2~pc is shown as the horizontal white, 
	labeled line.\label{vecmap_abcde}}
\end{figure}

The presence of multiple components of B-fields in RCW57A can be seen from 
the $H$-band polarization vector map overlaid on the $^{13}$CO(1~--~0) \citep{Purcelletal2009}
total integrated intensity image (Figure \ref{vecmap_abcde}) and 
the histogram of $\theta(H)$ (continuous histogram in Figure \ref{nirccd_and_ppahk_bffg}(d)). 
Each component follows its own spatial distribution and 
exhibits its own $\theta(H)$ distribution and dispersion. For this reason, we divided the observed area 
into five regions, named, A, B, C, D, and E, as shown in 
Figure \ref{vecmap_abcde} and Table \ref{mfstrength_cf_hilde}.
The spatial extent of each region is visually chosen in such a way that it should not 
include multiple components of $\theta(H)$. 
This criteria constrains the dispersion of $\theta(H)$ to be less 
than 25$\degr$ \citep{Ostrikeretal2001}, so as to apply the CF method to estimate the B-field strength. 
Some portions are not included because they either have polarization data but not CO data 
(the area right to the C region), or have CO data but has multiple 
$\theta(H)$ distributions (area slightly above the E region), 
or contains few randomly oriented vectors (area distributed between D and E regions). 
We believe that the bias involved in the selection of five regions may not have significant impact on 
the final scientific results. 

\begin{figure}[!ht]
\centering
\resizebox{7.5cm}{15cm}{\includegraphics{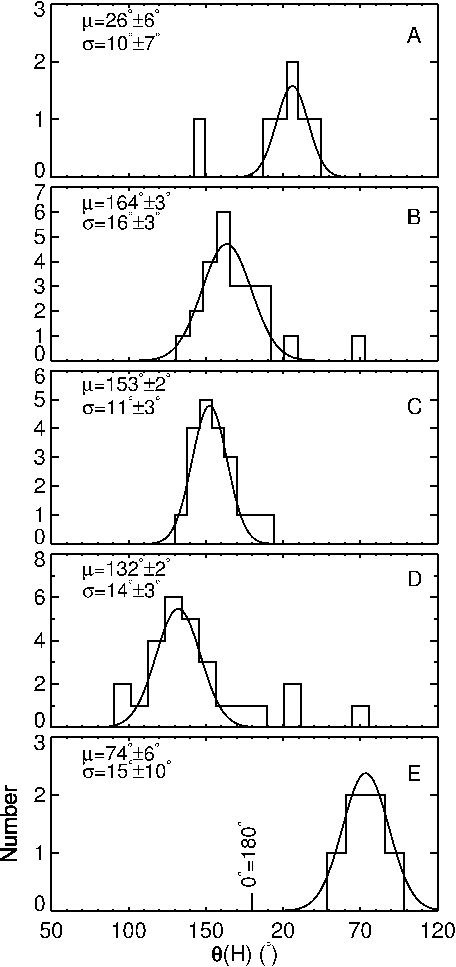}}
\caption{Frequency distribution of $\theta(H)$ for the regions A, B, C, D, and E. 
Gaussian fits are shown with black curves. Mean and uncertainties along with their fit errors are mentioned in each panel.
\label{distri_vel_pa_a}}
\end{figure}

A Gaussian function was fit to the $\theta(H)$ distribution for each region, 
as shown in Figure \ref{distri_vel_pa_a}. 
The fitted means and dispersions in polarization angles, 
along with their errors, are listed in columns 7 and 8 of Table \ref{mfstrength_cf_hilde}, respectively.
Mean $\theta(H)$ (=$\mu_{\theta(H)}$) values of regions A, B, C, D, and E regions 
found to be 26$\degr$, 164$\degr$, 153$\degr$, 
132$\degr$, and 74$\degr$ respectively. Dispersions in $\theta(H)$ (=$\sigma_{\theta(H)}$) 
of five regions lie between 10~--~16$\degr$. 

\subsubsection{\it Velocity dispersion: $\sigma_{V_{LSR}}$}\label{vlsrdeter}

\begin{figure}[!ht]
\centering
\resizebox{7.5cm}{15cm}{\includegraphics{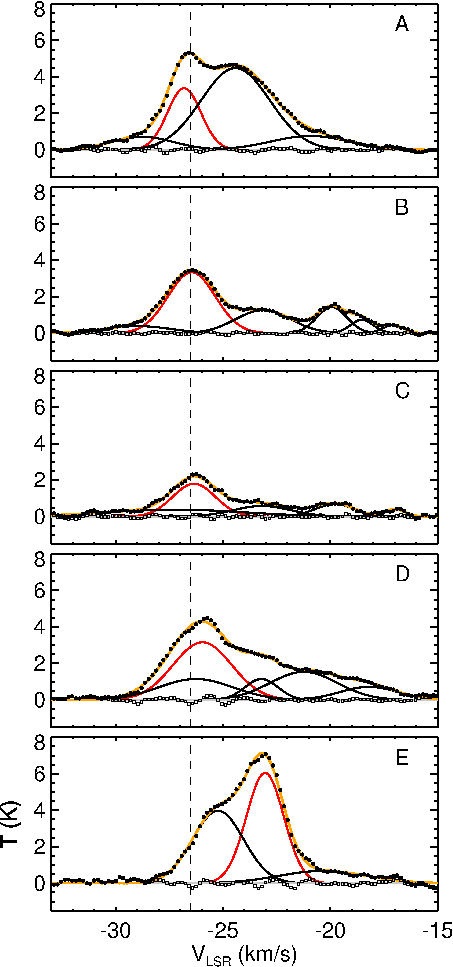}}
\caption{Mean $T (K)$ vs. $V_{LSR}$ (\kms) spectra for the 
regions A, B, C, D, and E (see Figure \ref{vecmap_abcde}) 
from $^{13}$CO(1~--~0) data of \citet{Purcelletal2009}. 
Observed data points are shown with filled dots. 
The spectrum of each region was fit with multiple Gaussians 
shown with black lines. The combined spectrum is shown with a thick orange line. 
The mean velocity corresponding to the RCW57A 
region is drawn at $-$26.5~\kms~using a dashed line.\label{distri_vel_pa_b}}
\end{figure}

To derive the velocity dispersion
in regions A, B, C, D and E, we have used 
$^{13}$CO(1~--~0) emission line mapping (Figure \ref{vecmap_abcde})
performed using the {\it Mopra} telescope by \citet{Purcelletal2009}. 
The observed $^{13}$CO(1~--~0) data has a velocity resolution of
0.4~\kms~ and a spatial resolution of 40$\arcsec$. 
CASA software was used to extract the mean velocity ($V_{LSR}$) versus 
brightness temperature ($T$) spectrum of each region, which are plotted in Figure \ref{distri_vel_pa_b} 
using small filled circles. 
All the regions exhibit at least two velocity components, 
with clear asymmetric spectra towards higher velocities. 
These asymmetric, spatially extended, spectra correspond to outflowing gas from the embedded YSOs, 
expanding I-fronts, or both. Each of these spectra ($T$ vs. $V_{LSR}$) 
were then fitted by a multi-Gaussian function using the custom IDL 
routine `gatorplot.pro'\footnote{\url{http://www.astro.ufl.edu/~warner/GatorPlot/}}. 
Details regarding this multi-Gaussian fitting 
are described in Appendix \ref{appendix_c}.

The resulting multi-Gaussian fitting parameters of each spectrum are presented in Table \ref{gausscomp_abcde} and are 
plotted with black lines in Figure \ref{distri_vel_pa_b}. 
The resulting combined spectrum of each region, the sum of all fitted multi-Gaussians (orange line), 
closely match the observed spectrum. The difference between observed and fitted spectra, the residuals, are 
shown with open squares and are closely distributed around zero $T$. 
Among the multiple velocity components of each region, 
we attribute the one related to the peak of the spectrum to the turbulent 
cloud component of that region (red Gaussian curve). These are used to estimate the B-field strength. 
The remaining Gaussian velocity components (black lines) are ascribed
to the expanding I-fronts and outflowing gas. 
These values are also given in column 9 of 
Table \ref{mfstrength_cf_hilde}. 

The dashed line drawn at $V_{LSR}$~$=$~$-26.5$~\kms~ (Figure \ref{distri_vel_pa_b}) corresponds to the line 
center velocity over the areas covered by ABCD regions of RCW57A. 
This $V_{LSR}$~$=$~$-26.5$~\kms~component closely
matches center of the red Gaussian components of all regions except for region E. 
However, based on the dense gas tracers (NH$_{3}$, N$_{2}$H$^{+}$, and CS), \citet{Purcelletal2009} 
witnessed that the peak $V_{LSR}$ of the filament remain constant
around  $-$24 \kms~(see their Figure 13), which is slightly different from
the $V_{LSR}$~$=$~$-26.5$~\kms. This is mainly because of the following reasons:
(a) CO is optically thick and hence traces the less dense outer parts of the cloud, 
and (b) the ABCD regions, located slightly away from the cloud center, 
contain less dense gas that is influenced by the H{\sc ii} region.

\begin{figure*}[!ht]
\centering
\resizebox{6.65cm}{7cm}{\includegraphics{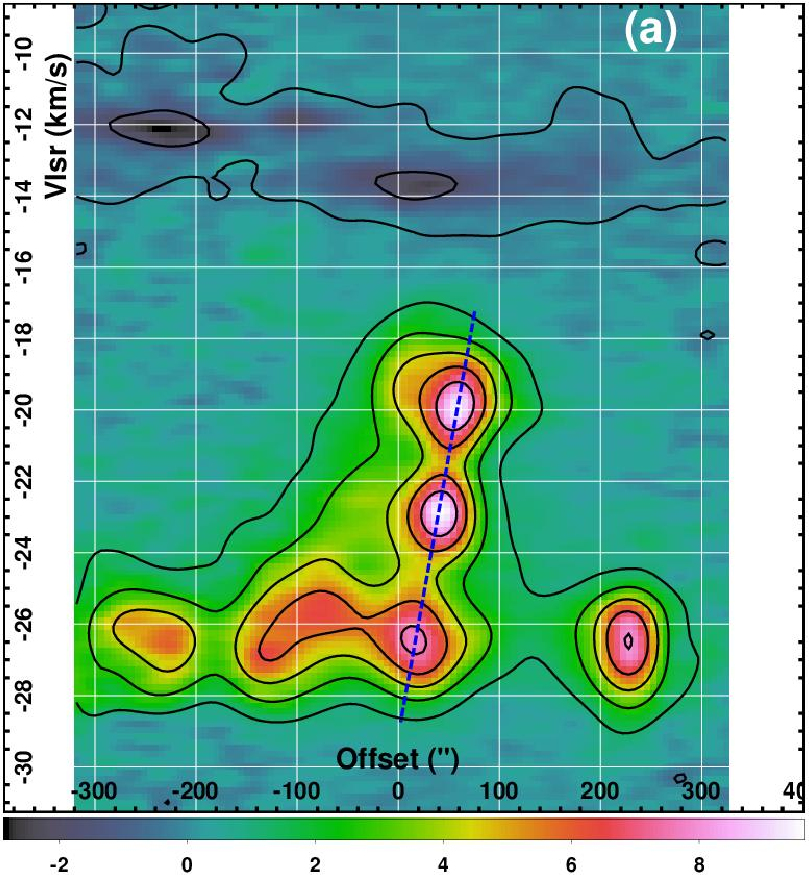}}
\resizebox{6.65cm}{7cm}{\includegraphics{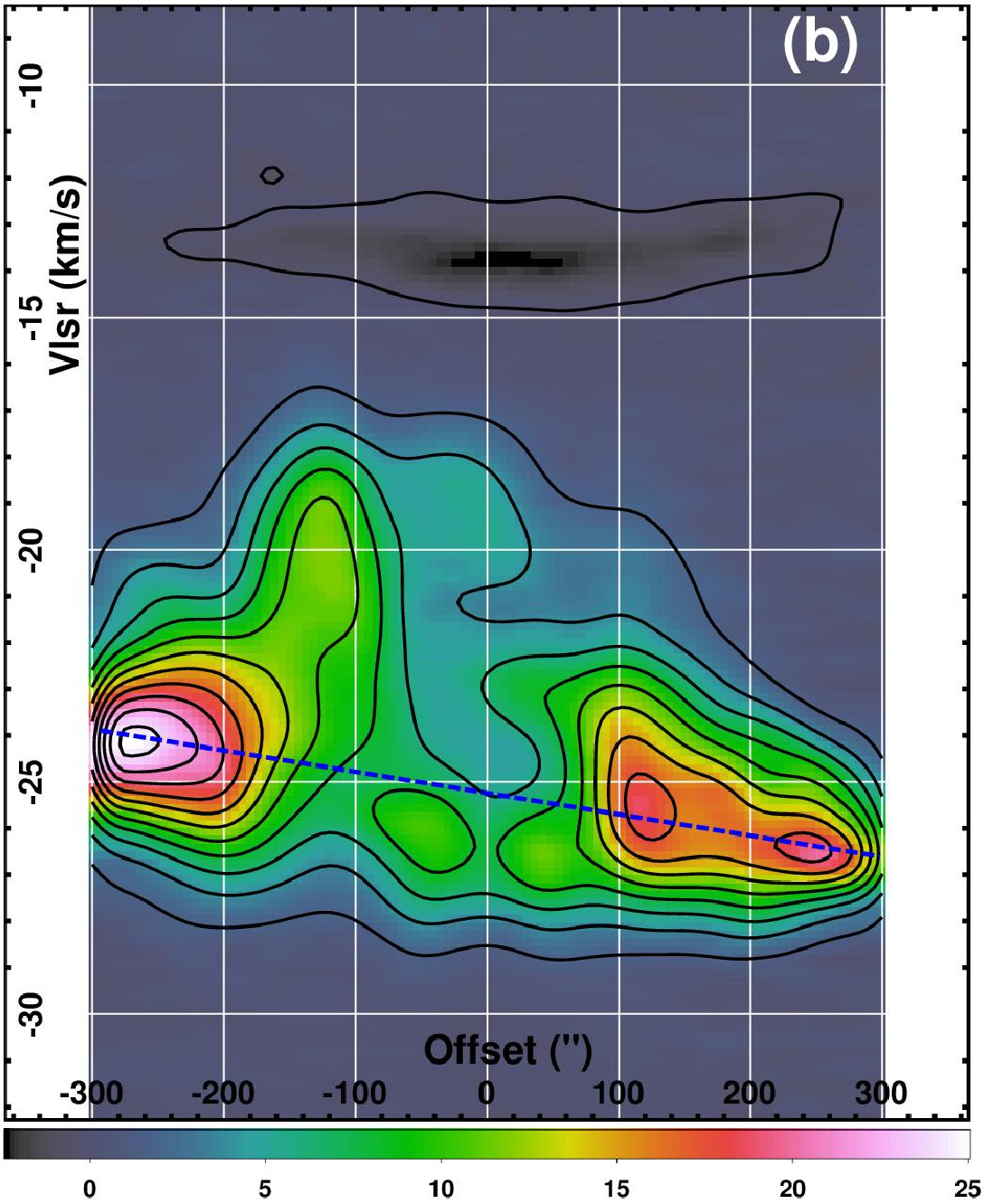}}
\caption{(a) Position-velocity diagram using a cut along the
probable orientation of outflowing gas (cut `1' with a position angle~$=$~163$\degr$, and
central offset coordinates are $\alpha$~$=$~$11^{\rmn h}12^{\rmn m}06^{\rmn s}$,
$\delta$~$=$~$-61\arcdeg18\arcmin56\arcsec$ [J2000]).
Contours are plotted at CO intensity ranging from $-$3 K to 10 K with a step of 1 K. Blue dotted line connects
the three brightest intensity points along the y-axis denoting a velocity gradient of
13 km~s$^{-1}$~pc$^{-1}$.
(b) Same as (a) but for a cut along the cloud major axis (cut `2' with a position angle~$=$~65$\degr$, and
central offset coordinates are $\alpha$~$=$~$11^{\rmn h}12^{\rmn m}04^{\rmn s}$,
$\delta$~$=$~$-61\arcdeg18\arcmin44\arcsec$ [J2000]).
Contours are plotted at CO intensity ranging from $-$2~K to 26~K with a step of 2~K. Blue dotted line connects
the brightest intensity points along the x-axis denoting a velocity gradient of
0.35~km~s$^{-1}$~pc$^{-1}$. Unit of the color scale in both panels is K.}
\label{pvcuts}
\end{figure*}

Below, we try to resolve and quantify
the contributions from turbulent cloud component, outflowing gas,
and expanding I-front using position-velocity (PV) diagrams. PV diagrams are useful diagnostics of gas kinematics in 
a star forming molecular clouds \citep[e.g., ][]{Purcelletal2009,ZhangCPetal2016}. 
Figures \ref{pvcuts}(a) and (b), show the $PV$ diagrams corresponding to the 
two cuts shown and labeled as `1' and `2', respectively, in Figure \ref{vecmap_abcde}. 
The cut `1' is along the orientation of expanding ionization front or 
outflowing gas (parallel to the minor axis of the filament) and 
the cut `2' is along the major axis of the filament. 
 
The velocity range over the whole region, based on the extent of the contours shown in Figures \ref{pvcuts}(a) and (b), 
spans $\simeq-$28.5~to $\simeq-$16.5~\kms~a width 
$\Delta V$~$\simeq$12~\kms. This implies that the entire region is dynamically active under the influence of 
H{\sc ii} region. However, among all parts of the cloud that we observed, one of the most quiescent clumps S4 \citep{Andreetal2008,Purcelletal2009} 
is situated to SE part of the filament. This clump is situated at $\sim$250$\arcsec$ and at $\sim-$26.5~\kms~ in Figure \ref{pvcuts}(b). 
Dust temperature $T_{d}$ ($\sim$19~--~24 K; \citealt{Andreetal2008}) and kinetic temperature $T_{k}$ ($\sim$15~--~25 K; \citealt{Purcelletal2009}) 
of the cores in S4, suggest that this clump is dynamically quiescent as being far from the H{\sc ii} region. 
Gas velocity around this clump spans over $\sim-$25~to $\sim-$28~\kms~ (centered at $\sim$-26.5~\kms) and a 
width of 2~--~3~\kms~(Figure \ref{pvcuts}(b)). Therefore, this velocity component at $\sim-$26.5~\kms~~is considered as the cloud $V_{LSR}$ component. 
Turbulent velocity dispersions of A, B, C, D, and E regions were also lie in the similar 
range of 2~--~3~\kms~as noted in Table \ref{gausscomp_abcde}. According to the collect and collapse 
model of triggered star formation \citep[see][and references therein]{Toriietal2015}, 
the velocity separation due to expanding ionization front based on the $PV$ 
diagrams to be $\sim$4~\kms. Therefore, the outflowing gas may have 
remaining contribution of $\sim$5~--~6~\kms~in addition to the cloud turbulent component of $\sim$2~--~3~\kms. 

Below we decompose the relative contribution of various velocity components in each region. 
In Figure \ref{pvcuts}(a), there exist 
two prominent red-shifted, high-velocity components with peak intensities
at $\simeq-$20~\kms~and $\simeq-$23~\kms. These components are concentrated close to the cloud center. 
While the former component is attributed to the 
outflowing gas, the latter one is to the expanding I-front. 
The signatures of both the components prevail in A, B, C, and D regions.  
However, as shown in the bottom panel of Figure \ref{distri_vel_pa_b}, the E region seems to be 
less influenced by the above two components, but has other two prominent components: 
(a) $\sim-$25~\kms~is related to the expanding I-front, and (b) 
$\sim-23$~\kms~is attributed to the turbulent cloud region. 
Moreover, spatially, the E region, in comparison to the others, is located far from the area
influenced by the outflowing gas as depicted with white arrows (cf. Figure \ref{polvecmap_bg}). 
This implies that the E region, containing a BRC, has a contribution
from the expanding I-fronts alone, while the rest of the regions had the influences of 
both expanding I-fronts and outflowing gas. 
Therefore, the gas velocity dispersion due to cloud turbulent component (red lines in Figure \ref{distri_vel_pa_b}) 
is well decomposed from other components using multi-Gaussian spectrum fitting.

\begin{figure*}[!ht]
\centering
\resizebox{7.25cm}{8.0cm}{\includegraphics{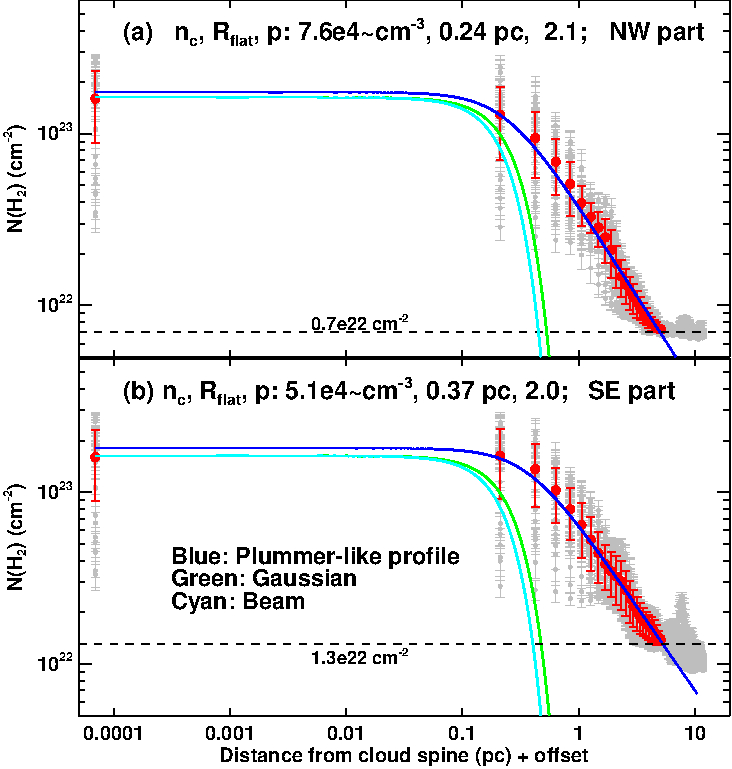}}
	\resizebox{8.25cm}{8.0cm}{\includegraphics{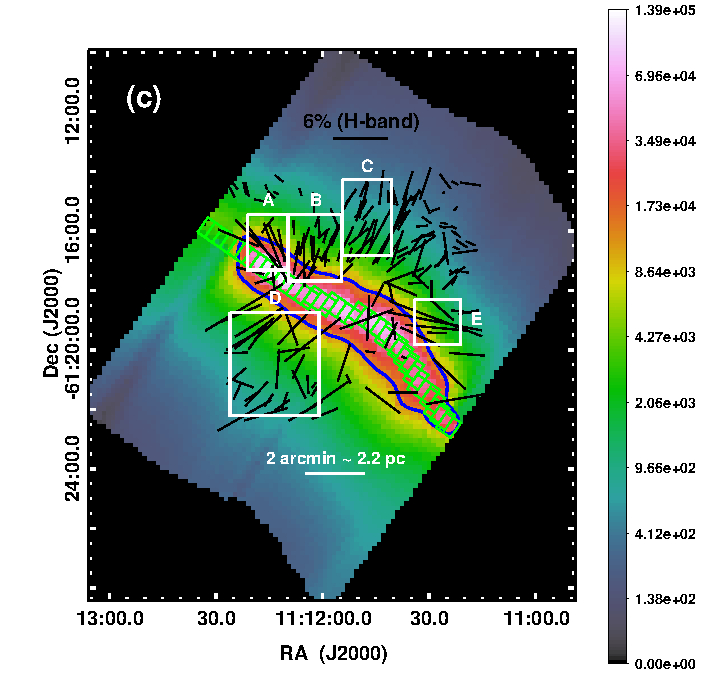}}
	\caption{(a) Plummer-like profile fitting (blue line) on the mean column density profile (red points) of the NW 
	region of RCW57A. Gray points corresponds to the individual NW column density profiles. Cyan and green 
         lines correspond to the 
	beam size (fwhm~$=$~$36\farcs4$~$=$~0.42~pc) of the volume density map and Gaussian fit (fwhm~=~0.50$\pm$0.18~pc), respectively. 
	(b) Same as (a) but for SE region of RCW57A. In both panels (a) and (b) the results from Plummer-like profile fitting 
	are mentioned. Dashed horizontal lines represent the background column densities in NE and SW regions. 
	(c) The volume density map constructed using Plummer-like profile fitting on the column density map.
        Spine points are shown with green squares with dimension of 36\farcs4$\times36\farcs4$ which are separated by 18.2$\arcsec$.
        The units of the color scale is cm$^{-3}$.  Thick blue contour is corresponding to the volume density $n$~$=$~13,000~cm$^{-3}$. Black vectors and white square boxes are same
        as those in Figure \ref{vecmap_abcde}. 
        A reference length scale with 2$\arcmin$ corresponding to 2.2~pc is shown.\label{voldensemap_abcde}}
\end{figure*}

\subsubsection{{\it Number density:}~$n(H_{2})$} 

{\it Herschel} PACS \citep{Poglitschetal2010} 160~$\mu$m and SPIRE \citep{Griffinetal2010}~250~$\mu$m, 
350~$\mu$m, and 500~$\mu$m images of RCW57A were used to construct a dust column density 
map by using the modified black body function (details are given in Appendix \ref{appendix_d}). 
The resolution of the final column density map is 36.4$\arcsec$. 
To obtain the number densities, we further used Plummer-like profile fitting on column density profiles 
to obtain the volume density map as described below.

To construct the column density ($N(H_{2})$) profiles, first we 
manually identify the cloud spine using the pixels that have maximum column densities and distributed along the filament.
Thus identified spine, depicted with green squares in Figure \ref{voldensemap_abcde}(c), divides the entire cloud region into two parts NW and SE. 
Second, for each spine element, we identify the $N(H_{2})$ profile elements distributed on a line that bisects the filament long axis 
and passes through that spine element. 
The dimensions of spine elements as well as $N(H_{2})$ profile elements are chosen as $36.4\arcsec\times36.4\arcsec$. 
To achieve the Nyquist sampling, separation between adjacent elements was chosen as 18.2$\arcsec$.
Third, we have constructed $N(H_{2})$ profiles as a function of projected distances from the
filament spine elements towards both NW and SE regions, and are plotted using gray circles in Figures \ref{voldensemap_abcde}(a) and (b), respectively. 
Every point and its uncertainty, in this plots, corresponds to the mean and standard deviation of column density 
values in a particular $N(H_{2})$ profile element. 
The mean column density profiles, using distance wise averaged column densities of the $N(H_{2})$ profile elements distributed parallel to the spine, 
are shown with red filled circles. Similarly, the error bars are corresponding to 
the standard deviations in corresponding mean column densities of $N(H_{2})$ profile elements. 

Assuming that the filament follows an idealized cylindrical model, we have extracted the volume density ($n(r)$) profile 
correspond to each $N(H_{2})$ profile  
by using Plummer-like profile \citep{Nutteretal2008,Arzoumanianetal2011,Juvelaetal2012,Palmeirimetal2013} 
fit to the $N(H_{2})$ profile 
	\begin{equation}\label{plumeq}
	n(r)=\frac{n_{c}}{[1+(r/R_{flat})^2]^{(p/2)}},
        \end{equation}

	\begin{equation}\label{plumeq2}
	N(r)~=~\frac{A_{p}n_{c}R_{flat}}{[1+(r/R_{flat})^2]^{(p-1)/2}}
	\end{equation}

The factor, $A_{p}$~=~$\pi$, is estimated using the equation 
$A_{p} = ~\int_{-\infty}^{+\infty}[(1+u^{2})^{p/2} \cos i]^{-1} du$ \citep{Arzoumanianetal2011} and by 
assuming that the filament is on the plane of the sky implying an inclination angle of $i$~$=$~0, and $p$~$=$~2. 
In equation \ref{plumeq2}, $N(r)$ is the column density profile as a function 
of offset distance ($r$) from the cloud spine. 
The fitting was performed using IDL {\sc mpfit} nonlinear least-squares fitting programme \citep{Markwardt2009} and 
extracted three parameters: (i) $n_{c}$ the central mean volume 
density of the spine point, (ii) $R_{flat}$ the radius within 
which the $N(H_{2})$ profile is flat, and (iii) $p$ the power-law index that determines the slope
of the power law fall beyond $R_{flat}$. 
Plummer-like profile fit was performed only to the data 
spanning to $\sim$5~pc (7.3 arcmin) from the spine, beyond which 
background column density dominates (horizontal dashed lines in Figures \ref{voldensemap_abcde}(a) \& (b)). 
The fitted values of $n_{c}$, $R_{flat}$, and $p$ are 
(7.6$\pm$1.3)$\times$10$^{4}$~cm$^{-3}$, 0.24$\pm$0.11~pc and 2.1$\pm$0.1 
for NW part (Figure \ref{voldensemap_abcde}(a)), 
and (5.1$\pm$0.9)$\times$10$^{4}$~cm$^{-3}$, 0.37$\pm$0.18~pc and 2.0$\pm$0.1 
for SE part (Figure \ref{voldensemap_abcde}(b)). 
These fit results, as indicated in the plots, are similar.

Volume density map, created using $n(r)$ profiles of all the filament spine elements, is shown in 
Figure \ref{voldensemap_abcde}(c). Since we considered spine point on the filament as the zeroth element 
while performing Plummer-like profile fitting for both the NW and SE regions, each spine point will have 
two volume density values and are averaged in the final map. The minimum and maximum $n(H_{2})$ values on the 
filament spine lie between 2.48$\times$10$^{3}$~cm$^{-3}$ and 1.52$\times$10$^{5}$~cm$^{-3}$.
The mean $n(H_{2})$ values for pixels at the extreme-end of the NW and SE regions are 111$\pm$26~cm$^{-3}$ and 
228$\pm$74~cm$^{-3}$, respectively. This implies that the RCW57A is still embedded in a diffuse background cloud. 

We found that more than 95\% of the polarization vectors are centered outside of
the blue isodensity contour corresponds to volume density $n(H_{2})$~$=$~13,000~cm$^{-3}$. 
It implies that our NIR polarimetry fails to trace adequately the highly 
extincted parts as as shown in Figure \ref{voldensemap_abcde}(c). 
Therefore, we exclude the pixels with $n$~$\geq$~13,000~cm$^{-3}$ while 
estimating the mean volume densities in A, B, C, D, and E regions. 
Means and Poisson errors (standard deviation divided by square 
root of the number of measurements) of the volume densities for the five regions 
(A, B, C, D, and E) are estimated and are listed in column 10 of Table \ref{mfstrength_cf_hilde}.
More details on producing volume density map from column density can also be found 
at \citet{Smithetal2014} and \citet{Hoqetal2017}.

It should be noted here that there exist 12 stars within the high column density regions (within the 
contour of volume density $n(H_{2})$~$=$~13,000~cm$^{-3}$ or 
column density $N(H_{2}) \simeq$1$\times$10$^{23}$~cm$^{-2}$). Only one star (star ID~=~290; Table \ref{polphotdata1}) is found to lie on 
the intrinsic locus of the late type stars in the NIR two-color diagram 
(with $[J-H]$~=~0.56$\pm$0.01 and $[H-K_{s}]$~=~0.19$\pm$0.01). The polarization characteristics 
($P(H)$~=~0.55$\pm$0.13 and $\theta(H)$~=~68.5$\pm$6.5) of this star are consistent with those of the FG stars. 
Although this star is classified as BG star 
based on its $[H-K_{s}]$~$>$~0.15, this star could be located close,  but foreground, to the cloud. Remaining 11 stars 
are consistent with either reddened cluster members embedded in the cloud or reddened field stars 
lying in the outskirts of the cloud but projected on the high density parts of the cloud region.

\subsection{\it Magnetic field strength using Chandrasekhar-Fermi method}\label{bstrength}

Using the dispersion in the polarization angles ($\sigma_{\theta(H)}$; column 8 of Table \ref{mfstrength_cf_hilde}), 
velocity dispersion values ($\sigma_{V_{LSR}}$; column 9 of Table \ref{mfstrength_cf_hilde}) 
and mean volume densities ($n(H_{2}$); column 10 of Table \ref{mfstrength_cf_hilde}), 
for the five regions of RCW57A, we estimated the B-field strength using the \citet{ChandrasekharFermi1953} relation:
\begin{equation}\label{cfrel}
B = Q~\sqrt{4\pi\rho}~\left(\frac{\sigma_{V_{LSR}}}{\sigma_{\theta_{H}}}\right). 
\end{equation}
The mass density $\rho$~$=$~$n(H_{2})$~$m_{H}$~$\mu_{H_{2}}$, where $n(H_{2})$ is the hydrogen volume density, $m_{H}$ is the mass of 
the hydrogen atom, and $\mu_{H_{2}}~\approx 2.8$ is the mean molecular weight per hydrogen molecule and includes the contribution from helium.
The correction factor $Q$~$=$~0.5 is included based on the studies using synthetic
polarization maps generated from numerically simulated clouds \citep{Ostrikeretal2001,Heitschetal2001} which suggest that for 
$\sigma_{\theta}$~$\leq$~25$\degr$, B-field strength is uncertain by a factor of two. 
Uncertainties in B-field strength were estimated by propagating 
the errors in $\sigma_{\theta(H)}$, $\sigma_{V_{LSR}}$ and $n(H_{2})$ values. 
The estimated B-field strengths and corresponding uncertainties are 
given in column 11 of Table \ref{mfstrength_cf_hilde}. 
The uncertainties in B-fields, derived by propagating errors 
in $n(H_{2})$, $\sigma_{V_{LSR}}$, and $\sigma_{\theta_{H}}$, are 
considerably large (see column 11 of Table \ref{mfstrength_cf_hilde}). 
We also estimate the B-field strengths and corresponding uncertainties without considering errors in $\sigma_{\theta_{H}}$ values. 
The resultant B-field strengths, given in column 12 of Table \ref{mfstrength_cf_hilde}, for A, B, C, D, and E regions are
to be 123$\pm$21$\mu$G, 89$\pm$13$\mu$G, 63$\pm$15$\mu$G, 108$\pm$27$\mu$G, and 74$\pm$8$\mu$G, respectively. 
Mean B-fields strength over five regions is estimated to be 91$\pm$8$\mu$G. 

To understand the importance of B-fields with respect to turbulence, we estimated 
the magnetic pressure and turbulent pressure using the 
relations $P_{B}$~$=$~$B^{2}/8\pi$ and $P_{\rmn turb}$~$=$~$\rho{\sigma_{\rmn turb}}^{2}$ 
(where $\sigma_{\rmn turb}$~$=$~$\sigma_{V_{LSR}}$ is given in Table \ref{mfstrength_cf_hilde}), respectively, 
and the results are given in Table \ref{mf_turb_pressure}. The mean $P_{B}$/$P_{\rmn turb}$ is estimated to be 2.4$\pm$0.6. 
These estimated parameters suggest that the 
magnetic pressure is more than the turbulent pressure at least by a factor of $\sim$2~in all five regions, 
signifying the dominant role of B-fields over turbulence. 
For the five regions together, the mean magnetic and turbulent pressures are 
estimated to be (35$\pm$7)$\times$10$^{-11}$ (dyn~cm$^{-2}$) and
(15$\pm$3)$\times$10$^{-11}$ (dyn~cm$^{-2}$), respectively. 
\citet{Danziger1974} 
estimated the mean electron density, $n_{e}$, as $\sim$20~cm$^{-3}$, and mean electron temperature, $T_{e}$, 
as $\sim$10000 K for the RCW57A region. For the entire region the thermal pressure $P_{th}$ 
using the relation $P_{th}$~$\simeq$~$2n_{e}kT_{e}$ (where $k$ is the Boltzmann constant), 
is estimated to be 6$\times$10$^{-11}$~dyn~cm$^{-2}$, which is smaller than the mean magnetic pressure. 
Ratio of mean magnetic to thermal pressure, $P_{B}/P_{th}$, is found to be $\sim$6.
A lower value of $P_{th}$ again suggests dominant magnetic pressure over thermal pressure. 
Therefore, in comparison to thermal and turbulent pressures, 
magnetic pressure is playing a crucial role 
in guiding the expanding ionization fronts or outflowing gas in RCW57A.  
The implications on the relative importance of B-field pressure in comparison to turbulent and thermal pressures 
on the formation and evolution of bipolar bubble is discussed in the Section \ref{sec_schematic}. 

\section{\sc Discussion}\label{sec_discuss}

\subsection{\it Anisotropic distribution of material in the cloud and the bipolar bubble}
Based on the CO and {\it Spitzer} images, number of stars with polarization detection, and the column 
density profiles we infer on the distribution of gas and dust in RCW57A. 
The quiescent molecular cloud gas traced by CO (red background in Figure \ref{polvecmap_bg}) 
is absent at the foot points of, as well as along, the bipolar bubbles. Moreover, the spatial distributions of the CO gas and the bubbles
(green background by {\it Spitzer} images) appear to be anti coincident with each other. 
This implies that cloud material has been either eroded or blown out by the expanding I-fronts, outflowing gas, or strong stellar winds from
the embedded early type protostars. Therefore, CO and {\it Spitzer} maps suggest that bipolar bubble likely blown by
the massive stars emanated stellar winds, outflowing gas, and expanding ionization fronts \citep[see ][]{Townsley2009}.

While the SE bubble appears to be closed towards the SE, the NW bubble is widely extended
towards the NW (Figures \ref{WISE4p6_K_R_color_composite} and \ref{polvecmap_bg}) and exhibits
large-scale loops, likely due to anisotropic distribution of material towards SE and NW regions. 
This is further corroborated by the following facts.
Although the area together covered by regions A, B and C (located in NW) is similar to that of D (located in SE),
the total number of background stars with polarization detection (51) in A, B, and C exceeds those in D (21) (see Table \ref{mfstrength_cf_hilde},
and Figures \ref{polvecmap_bg} and \ref{vecmap_abcde}). 
Furthermore, the mean $N(H_{2})$ profiles suggests that anisotropic distribution of gas and dust in RCW57A i.e., SE part
($N(H_{2})$~$\sim$1.3$\times$10$^{22}$ cm$^{-2}$) has relatively
more background material than in the NW ($N(H_{2})$~$\sim$0.70$\times$10$^{22}$ cm$^{-2}$) as depicted with dashed lines
in Figures \ref{voldensemap_abcde}(a) and (b). Similar asymmetric distribution of column density profiles
was seen towards other region \citep[e.g., Taurus B211/3 filament;][]{Palmeirimetal2013}.

An X-ray study \citep[][see their Figures 3 and 4]{Townsleyetal2011a} revealed the presence of OB stellar association (NGC\,3576OB) 
towards NW of the main embedded star forming region RCW57A (or NGC\,3576GHIIR). Furthermore, presence of hard X-ray emission and 
a pulsar (+PSRJ1112-6103 \& PWN) in NGC\,3576OB region may suggests a fast occurrence of supernovae event. 
This supernova together with OB stars of NGC\,3576OB stellar association, most probably, evacuated the material in the NW part of the RCW57A. 
This could be the plausible reason behind the low density material observed in the NW region. 
The observed large scale loops in the NE, possibly, formed due to the newly formed material as 
a result of expanding I-fronts driven by the embedded massive protostars of RCW57A. 

A high velocity outflow signature discernible close to the eastward from the center, while the same feature is 
absent to the westward as shown in Figure \ref{pvcuts}(b). This could attribute to the confined ionized gas flow 
towards west from center possibly due to the presence of a dense clump (black thin contours in Figure \ref{polvecmap_bg}). 
High velocity component is only seen towards the east which implies dearth of
dense material there and hence ionized gas expanded with relatively more velocity.
\citet{dePreeetal1999} and \citet{Purcelletal2009} have also shown that the H{\sc ii} region 
expands more freely towards the east and is bounded to the west.

\subsection{\it Column density profiles: filament width and B-field support}\label{sec_cold_discuss}

We have fitted a Gaussian to the mean column density profiles spread over $\sim$5~pc radii and resultant 
FWHM, corresponds to the filament characteristic width, is found to be
~1.55$\pm$0.09~pc. However, a constant filament width $\sim$0.1~pc has been 
determined for the central dense sections of filaments of the nearby star forming regions
in the Herschel Gould Belt survey by fitting a Gaussian to the inner sections (radii upto 0.3~--~0.4~pc) 
of the filaments by \citep{Arzoumanianetal2011}. Therefore, we fit a Gaussian to the inner 1~pc width of the clubbed mean column density
profiles of NW and SE regions and the resultant $\sigma$ and FWHM 
are found to be 0.21$\pm$0.08~pc 0.50$\pm$0.18~pc, respectively.
Within the error, the filament width (0.50~pc; green lines in Figures \ref{voldensemap_abcde}(a) and (b))
of RCW57A is higher than the constant filament width $\sim$0.1~pc \citep{Arzoumanianetal2011}.
Filament widths larger than 0.1~pc~(spanning over 0.1~pc~--~1~pc with the typical widths of $\sim$0.2~--~0.3~pc) 
has also been witnessed in a number of studies \citep[eg.,][]{Hennemannetal2012,Juvelaetal2012}. 
However, the filament width of RCW57A is not resolved because of the limited resolution
(36$\farcs$4 corresponds to 0.42\,pc as shown with cyan lines) of the column density map.

The mean power-law index $p$~=~2.02$\pm$0.05, derived from Plummer-like profile fitting for RCW57A, 
is consistent with other studies \citep{Arzoumanianetal2011,Juvelaetal2012,Palmeirimetal2013,Hoqetal2017}.
This similar value of index ($p$~=~2) is expected if the filament is supported by
B-fields \citep{Hennebelle2003,TilleyPudritz2003} which is
in accordance with our observational results supporting active role of B-fields
in the formation and evolution of filament as described in Section \ref{sec_schematic} below.

\subsection{\it Evolutionary scenario of filament and bipolar bubble}\label{sec_schematic}

\begin{figure*}[!ht]
\centering
\resizebox{14cm}{12cm}{\includegraphics{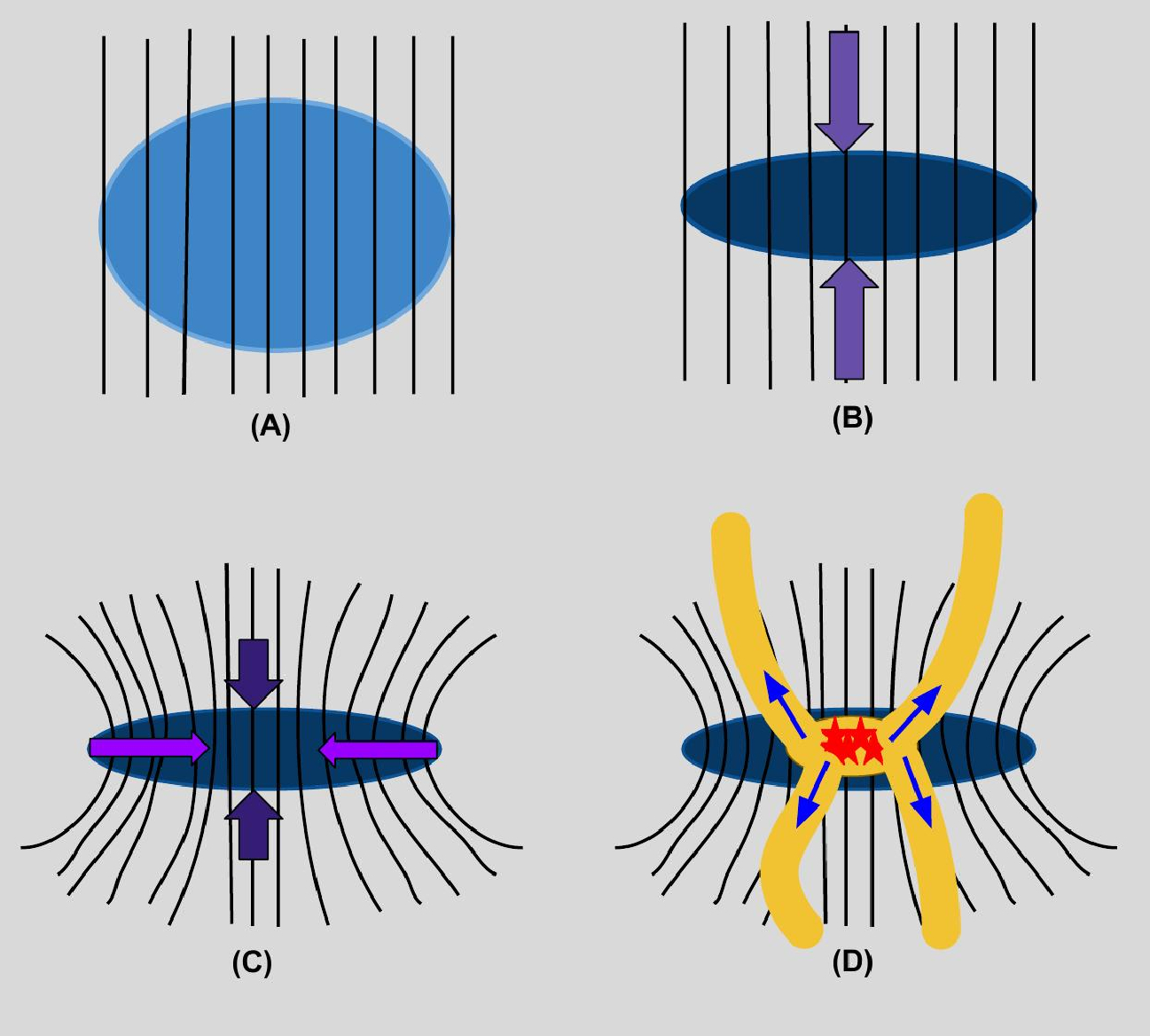}}
\caption{Schematic diagram illustrates the possible connection between filament, bipolar bubble and B-fields. 
Initially, subcritical molecular cloud permeated by uniform B-fields (schema A) could have gravitationally 
compressed along the B-fields into a filament (schema B). 
Later, as mass accumulated at the center of the filament, gravitationally contracting cloud could have pulled 
the B-fields along the cloud material towards the central region. 
Thus, B-fields followed a clear hour-glass morphology when the collapsing cloud reached supercritical state (schema C). 
Dense and high pressurized H{\sc ii} region formed due to the formation of massive stars at the cloud center (schema D). 
Further, propagation of ionization fronts and stellar winds is rather controlled by the 
an isotropic distribution of gas and dust in the filament. Presence of B-fields in an hour-glass morphology is 
further introduced an additional anistropic pressure in favor of expanding ionization fronts or outflowing gas so as 
to form the bipolar bubble (schema D). 
}
\label{schematic}
\end{figure*}

Based on the morphological correlations among (i) dense filamentary cloud structure seen in sub-mm (at 0.87-mm ATLASGAL and 1.1-mm SIMBA
dust continuum emission maps), (ii) bipolar bubble seen in mid infrared images, and (iii) the B-field morphology of RCW57A trace by NIR polarimetry, 
we postulate a scenario as illustrated with four schemas in Figure \ref{schematic}. 
Each schema describes 
our observational evidence along with our prediction at a particular evolutionary state of RCW57A.

{\bf Schemas A and B:} As per the background star polarimetry, the filament (position angle $\sim60\degr$; Figure \ref{polvecmap_bg}) 
is oriented perpendicular to the B-fields (the dominant component of $\theta(H)$~=~$\sim163\degr$; Figure \ref{nirccd_and_ppahk_bffg}(d)). 
From this observational evidence, we speculate the formation history of filament and the role of B-fields in them. 
Initially a subcritical molecular cloud is supported by B-fields as shown in schema A, later 
compresses into a filamentary molecular cloud due to B-fields guided gravitational contraction as per schema B. 

Although the role of B-fields in the formation and evolution of
filamentary molecular clouds is still a matter of debate, its
morphology with respect to the geometry of filaments is well established.
The geometry of B-fields in a molecular cloud is mainly governed by
the relative dynamical importance of magnetic forces to gravity and turbulence.

If the B-fields are dynamically unimportant compared
to the turbulence, the random motions dominate the structural dynamics
of the clouds, and the field lines would be dragged along the turbulent
eddies \citep{Ballesteros-Paredesetal1999}. In that case, B-fields would
exhibit chaotic or disturbed B-field structures. 
Based on the facts that the B-field structure is more regular with an hour-glass morphology and B-field pressure dominates over turbulent pressure 
(see Section \ref{bstrength}), we do not favor weak B-field scenario in RCW57A. 

If B-fields are dynamically important
then the support to the molecular cloud against gravity is
rendered predominantly by the B-fields. In this case, B-fields would be aligned, preferentially, perpendicular
to the major axis of the cloud \citep{Mouschovias1978}. This is expected as the cloud tends to contract more in the direction
parallel to the B-fields than in the direction perpendicular to the it.
Recent polarization studies have also demonstrated that B-fields
are well ordered near dense filaments and perpendicular to their 
long axis \citep{Monetietal1984,PereyraMagalhaes2004,Alvesetal2008,Chapmanetal2011,Sugitanietal2011,Palmeirimetal2013,Lietal2013,FrancoAlves2015,Coxetal2016,PlanckCollaborationXXXII2016}. 
Numerical simulations have also shown that filamentary molecular clouds form by gravitational compression 
guided by B-fields \citep[e.g., ][]{NakamuraLi2008,Lietal2013}. Therefore, we believe that 
the filament in RCW57A could have formed due to
the gravitational compression of cloud along the B-fields. 

{\bf Schema C}: Due to the accumulation of more mass at the center, the filament is achieved a supercritical state 
from its initial B-field supported subcritical state. 
B-fields are dragged along the gravity-driven material contraction towards the strong 
gravitational potential at the cloud center, and subsequently they configured into an hour-glass morphology (schema C). 
Similar hour-glass morphology of B-fields were observed towards Orion molecular cloud 1 \citep{Schleuning1998,Vaillancourtetal2008,WardThompsonetal2017} 
and Serpens cloud core \citep{Sugitanietal2010} signifying ongoing gravitational collapse.  

In RCW57A, B-fields in some portions exhibit clear evidence for hour-glass morphology. 
In the NE part of the filament (covered by two regions A and D; see Figure \ref{vecmap_abcde}),
B-fields are bent by $\sim90^\circ$ as the mean $\theta(H)$ for region A is $\sim$26$\degr$ and 
for D is $\sim$132$\degr$ (see column 7 of Table \ref{mfstrength_cf_hilde}). 
Therefore, B-fields are distorted in the NE part of the filament, signifying a
collapsing filamentary molecular cloud. However, this similar distortion is not clearly 
apparent in the SW part of the filament,
partly because of the light from the background stars is obscured by the dense cloud material. 
Therefore, the NIR polarimetric observations were unable to detect 
sufficient numbers of background stars in the SW part to
delineate the detailed structure of the B-field there. 

{\bf Schema D}: Gravitational instability caused the filament fragmentation into seven cores (blue diamonds in Figure \ref{polvecmap_bg}) 
as evident from dust continuum emission maps \citep{Andreetal2008} as well as dense gas tracers \citep{Purcelletal2009}. 
Star formation has began at the central region due to the onset of gravitational collapse during which B-fields were not strong enough to 
stop the cloud collapse. 
Ionization and shock fronts, driven by the H{\sc ii} region, are propagated in to the ambient medium. 
Interaction between the H{\sc ii} region and dense cloud material 
is evident from Figure \ref{polvecmap_bg} and 
as a consequence of this, several IRS sources are formed 
at the boundary of interaction 
signifying triggered star formation at the edges of H{\sc ii} region of RCW\,57A. 

Our observational results suggest that B-field pressure dominates over thermal pressure (see Section \ref{bstrength}), 
implying active role of B-fields in governing the feedback processes. 
Since the B-fields observed to be responsible for the formation of filament in RCW57A (Schema B), 
it is expected that B-fields anchored through the filament 
may also have significant influence on the expanding I-fronts. 
In addition to the anisotropic pressure that the I-fronts experience while they expand anisotropically in the filament, they 
undergo an additional anisotropic pressure in the presence of strong B-fields \citep{Bisnovatyi-Kogan1995}. 
As a result I-fronts experience accelerated flows along the B-fields (i.e., low pressure) 
and hindered flows (i.e., high pressure) in the perpendicular direction 
to the B-fields \citep{Tomisaka1992,Gaensler1998,PavelClemens2012,vanMarleetal2015}. 
Therefore, I-fronts expand into greater extent along the hour-glass shaped 
B-fields and eventually form the bipolar bubble as shown in schema D of Figure \ref{schematic}.

Similar applications for the influence of B-fields on I-fronts have found at other environments. 
Bubbles around young H{\sc ii} regions and supernova remnants have been found to be elongated along the 
Galactic B-field orientation \citep{Tomisaka1992,Gaensler1998,PavelClemens2012}.  
\citet{FalcetaGoncalvesetal2014} have studied that the strong B-fields can lead to form the bipolar planetary nebula. 
\citet{vanMarleetal2015} have investigated the influence of B-fields on the expanding circumstellar
bubble around a massive star using magneto-hydrodynamical simulations. 
They found that 
that the weak B-fields cause circumstellar bubbles become ovoid rather than spherical shape. 
On the other hand, strong B-fields lead to the formation of a tube-like bubble. 

Therefore, according to the proposed evolutionary scenario (Figure \ref{schematic}), initially B-fields in molecular clouds may be 
important in guiding the gravitational compression of cloud material to form a filament. In the later stage, hour-glass B-fields might have 
guided the expansion and propagation of I-fronts to form the bipolar bubble. 

\subsection{\it B-fields favoring the formation of massive cluster at the center of the filament}\label{sec_cond_favour2}

Based on the presence of 29 NIR excess sources \citep{Persietal1994} and finding of 51 stars earlier than A0 
\citep{Maerckeretal2006}, it has been found that RCW57A hosts a massive infrared cluster at the center 
(within the cyan contour encloses H{\sc ii} region in Figure \ref{polvecmap_bg}) of the filament. 
Furthermore, the {\it Chandra} X-ray survey by \citet[][see also \citealt{Townsleyetal2011a}]{Townsleyetal2014} has provided a first 
detailed census about the members in the embedded
massive young stellar cluster (MYSC; see their Figures 9c and 9d). Their study unraveled
many highly-obscured but luminous X-ray sources (one source exhibits photon pile-up)
that are likely the cluster's massive stars ionizing its H{\sc ii} region.
This highly-obscured X-ray cluster is situated at the SW edge of the infrared cluster. 
Formation history of this cluster is poorly understood. Below, we propose that the large scale B-fields traced by NIR polarimetry may also
govern crucial role in the formation of this massive cluster.

Initial turbulence in the molecular clouds may decay rapidly. To maintain turbulence in the cloud and hence 
to govern cloud stability against rapid gravitational collapse, supersonic turbulence should be 
replenished by means of prostellar outflows. In the model of outflow driven turbulence cluster formation 
\citep[][and references therein]{LiNakamura2006,NakamuraLi2007,NakamuraLi2011}, 
B-fields are dynamically important in governing two processes. First, the outflow energy and momentum are
allowed by the B-fields to escape from the central star forming region into the ambient cloud medium.
Second, B-fields guide the gravitational infalling material (just outside of the outflow zone) to the central region.
Therefore, the competition between protostellar outflow-driven turbulence and gravitational infall regulates
the formation of stellar clusters in the presence of 
dynamically important B-fields. Observational signatures towards the Serpens cloud core 
\citep{Sugitanietal2010} 
is in accordance with the scenario of B-field regulated cluster formation: 
(a) B-fields should align with the short axis of the filament, outflows and gravitational infall motions, and (b) 
outflow injected energy should be more than the dissipated turbulence energy in order to maintain the supersonic turbulence in the cloud. 
In the case of RCW57A, the presence of a deeply embedded cluster, and 
alignment of the I-fronts, outflowing gas, and bipolar bubbles with the NIR-traced B-fields, 
trace the pre-existing conditions in favor of cluster formation according to the 
protostellar outflow driven turbulence cluster formation in the presence of dynamically strong B-fields.

\section{\sc Summary and Conclusions}\label{conclusions}
Though there exist few studies regarding 
the formation of bipolar bubbles, none of them has explored the importance of B-fields. 
Our aim was to understand the morphological correlations among the B-fields, filament, and bipolar bubbles, 
and their implications for the star formation history in RCW57A. We conducted NIR polarimetric observations in the $JHK_{s}$-bands using SIRPOL to delineate the B-field structure in RCW57A. 
We employed various means (NIR and MIR color-color diagrams, polarization efficiency diagrams, and the polarization 
characteristics) to exclude YSOs with possible intrinsic polarization. 
Through our analyses, we separated 97 confirmed foreground stars from 178 confirmed background stars having $H$-band polarimetry. 
Below we summarize the results of our present work. 

\begin{itemize}

\item The foreground dust 
dominated by a single component B-field having a mean $\theta(H)$ $\sim65\degr$, which is 
different from the position angle (101$\degr$) corresponding to the Galactic plane.  

\item The polarization values of the background stars consistent with the dichroic origin of dust polarization and 
exhibit relatively higher polarization efficiencies, suggesting efficiently aligned dust grains in RCW57A. 

\item The polarization angles of the reddened background stars reveal that the B-field in RCW57A is configured into an hour-glass 
morphology which follows closely the structure of the bipolar bubbles. The dominant component of the B-field ($\theta(H)\sim 163\degr$) is perpendicular to the 
filament major axis. The orientations of both outflowing gas from the embedded protostars and the expanding ionization
fronts from the H{\sc ii} region are aligned with the B-fields. 

\item The B-field strength averaged over five regions across RCW57A based on the Chandrasekhar-Fermi method, 
is 91$\pm$8$\mu$G. The B-field pressure is estimated to be more than turbulent and thermal pressures. 

\item Morphological correlations among B-fields, filament, and bipolar bubbles as well as the 
dominance of B-field pressure over turbulent and thermal pressures suggest a scenario in which 
B-fields not only play an important role in formation the filamentary molecular cloud
but also in guiding the expansion and propagation of I-fronts  and/or outflowing gas to 
form bipolar bubbles. In this picture, B-fields impart additional anisotropic pressure to 
expanding I-fronts, from H{\sc ii} regions, in order to be expanded and propagated into greater extent. 

\item Protostellar outflow driven turbulence and gravity in the presence of dynamically important B-fields might be
responsible for the cluster formation. Therefore, our study, based on the link between B-fields, filament, and bipolar bubble, traces the preexisting conditions where B-fields might also be important in the formation of massive cluster in RCW57A.
 
\end{itemize}

\section*{Acknowledgments}

We thank Prof D. P. Clemens for helpful discussions and providing constructive 
suggestions in the improvement of the draft. We thank the anonymous referee for his/her careful reading 
and insightful comments which have improved the contents of the paper. 
CE, SPL and JWW are thankful to the support from 
the Ministry of Science and Technology (MoST) of Taiwan through grants 
102-2119-M-007-004-MY3, 105-2119-M-007-022-MY3, and 
105-2119-M-007-024. CE and WPC acknowledge the financial support from the 
grant 103-2112-M-008-024-MY3 funded by MoST. MT is partly supported by JSPS KAKENHI Grant (15H02036).
AMM and his team's activities at IAG-USP are partially supported by grants from the 
Brazilian agencies CNPq, CAPES and FAPESP. CE thank Dr M. R. Samal for fruitful discussons. 
This publication makes use of data from the 2MASS (a joint project of the
University of Massachusetts and the Infrared Processing and Analysis Center
/California Institute of Technology, funded by the National Aeronautics
and Space Administration and the National Science Foundation). \\

{\it{Facility}}: IRSF (imaging polarimeter SIRPOL)


\appendix
{\Large \bf Appendix} 

\section{\it Calibration of IRSF photometry into {\rmn 2MASS}}\label{appendix_a}

The 401 stars having both IRSF and 2MASS data were used to calibrate the IRSF photometry of 702 stars into the 2MASS system 
using the following relations:

\begin{eqnarray}\label{irsfto2masstrans}
{J_{\rmn 2MASS} = J_{\rmn IRSF} + \alpha_{1} \times \left[J-H\right]_{\rmn IRSF} + \beta_{1}} \\
{H_{\rmn 2MASS} = H_{\rmn IRSF} + \alpha_{2} \times \left[H-K_{s}\right]_{\rmn IRSF} + \beta_{2}} \\
{{K_{s}}_{~\rmn 2MASS} = {K_{s}}_{~\rmn IRSF} + \alpha_{3} \times {\left[J-K_{s}\right]}_{\rmn IRSF} + \beta_{3}} \\
{\left[J-H\right]_{~\rmn 2MASS} = \alpha_{4} \times \left[J-H\right]_{\rmn IRSF} + \beta_{4}} \\
{{\left[H-K_{s}\right]}_{\rmn 2MASS} = \alpha_{5} \times {\left[H-K_{s}\right]}_{\rmn IRSF} + \beta_{5}} 
\end{eqnarray} 

\begin{figure}[!ht]
\centering
\resizebox{7cm}{17.0cm}{\includegraphics{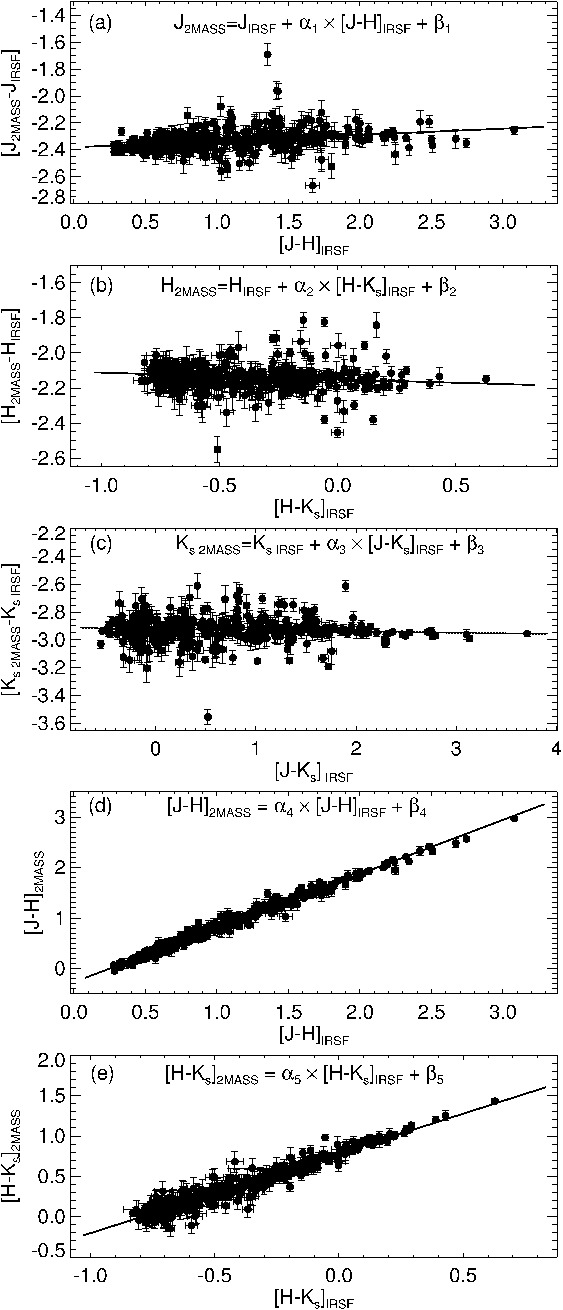}}
\caption{Calibration between IRSF and 2MASS systems by using $JHK_{s}$ photometric magnitudes 
of 401 stars whose photometric 
uncertainties are less than 0.1 mag. (a) $[J-H]_{\rmn IRSF}$ vs. $[J_{\rmn 2MASS}-J_{\rmn IRSF}]$ 
(b) ${[H-K_{s}]}_{\rmn IRSF}$ vs. $[H_{\rmn 2MASS}-H_{\rmn IRSF}]$, 
(c) ${[J-K_{s}]}_{\rmn IRSF}$ vs. $[{K_{s}}_{~\rmn 2MASS}-{K_{s}}_{~\rmn IRSF}]$ 
(d) $[J-H]_{\rmn IRSF}$ vs. $[J-H]_{~\rmn 2MASS}$, and (e) ${[H-K_{s}]}_{\rmn IRSF}$ vs. ${[H-K_{s}]}_{\rmn 2MASS}$. \label{2masstosirius}}
\end{figure}

Figure \ref{2masstosirius} depicts the weighted least square fittings for magnitudes and colors between IRSF and 2MASS. 
The fitted coefficients are found to be: 
$\alpha_{1}=$0.0462$\pm$0.0033, $\beta_{1}=-$2.3812$\pm$0.0040; 
$\alpha_{2}=-$0.0384$\pm$0.0053, $\beta_{2}=-$2.1506$\pm$0.0028;
$\alpha_{3}=-$0.0095$\pm$0.0018, $\beta_{3}=-$2.9199$\pm$0.0021;
$\alpha_{4}=$1.0719$\pm$0.0051, $\beta_{4}=-$0.2696$\pm$0.0060; and 
$\alpha_{5}=$0.9872$\pm$0.0078, $\beta_{5}=$0.7828$\pm$0.0042. 
These coefficients were used, finally, to transform the 
instrumental magnitudes and colors of the IRSF system to the 2MASS system. 

\section{\it B-field structure around a BRC}\label{appendix_b}
\begin{figure*}[!ht]
\centering
\resizebox{9cm}{9cm}{\includegraphics{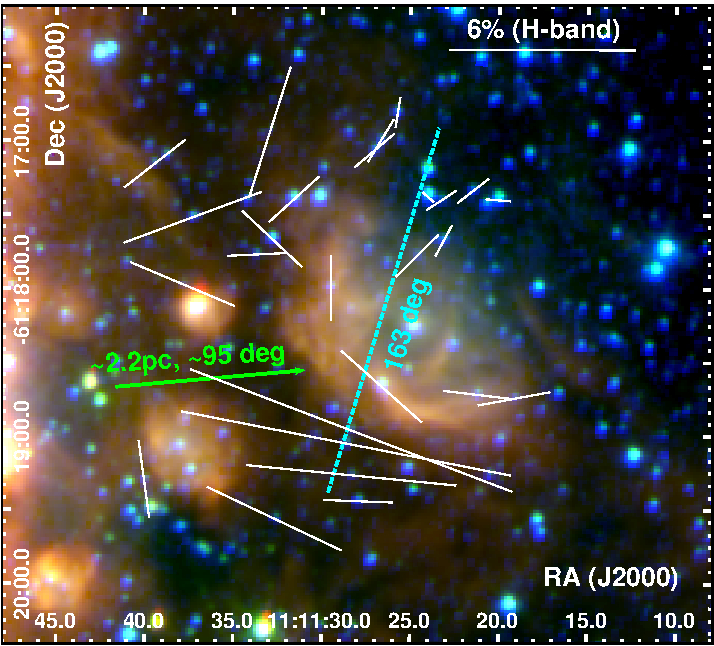}}
\caption{Polarization vectors (white) depict the B-field orientation, using the $H$-band polarimetry of 26 confirmed BG 
stars, around a BRC located within a white
square box in Figure \ref{polvecmap_bg}.
Background is the tricolor image ($\sim4.5\arcmin\times4.3\arcmin$) made using
{\it Spitzer} IRAC Ch4 (red), IRAC Ch2+Ch3 (green) and
2MASS $K_{s}$-band (blue) images.
Green arrow with an orientation of $\sim95\degr$ indicates the direction of ionizing front emanated from
H{\sc ii} region. Dashed cyan line denotes the B-field orientation ($\sim 163\degr$)
in the star-forming region prior to the formation of the BRC. Reference vector with 6\% is shown.}
\label{Bfield_around_BRC}
\end{figure*}

The $H$-band polarimetry of 26 background stars covering the BRC depicts compressed B-fields along the rim of the BRC 
as shown in Figure \ref{Bfield_around_BRC}. 
BRCs are formed in and around H{\sc ii} regions due to interaction of I-fronts 
with the ambient medium. It is unclear whether the pre-existing inhomogeneities in the medium (in the form of globules)
or instability in the expanding I-fronts are important in the formation of 
these structures \citep{Kahn1958,Williamsetal2001}, 
but both might be acting albeit on different length scales \citep{MackeyLim2013}. Though thermal pressure enhanced by
the ionizing radiation governs the entire formation processes of BRCs, it has been shown that
B-field pressure and its orientation may also play a key role in shaping the density distribution of the BRCs 
as well as their structure \citep{Motoyamaetal2013}. \citet{Motoyamaetal2013} also stated that the evolution of the BRCs 
depends on the orientation of B-fields in reference with the direction of the exciting source(s) (or I-front).

\citet{MackeyLim2011MNRAS,MackeyLim2011BSRSL,Mackey2012,MackeyLim2013} 
have performed 3D radiation-magnetohydrodynamic (RMHD) simulations on photoionization of dense clouds threaded by
B-fields with different orientations and strengths with respect to the direction of ionizing source.
Here we correlate our results, especially with those models involving different B-field strengths but oriented perpendicular to the
direction of propagating I-fronts. The B-fields in the models involving weak ($\sim$15~$\mu$G) and moderate ($\sim$50~$\mu$G) strengths 
are swept into alignment with the pillar/globule during the later evolutionary stages 
\citep[][see also \citealt{Henneyetal2009}]{MackeyLim2011MNRAS}, whereas at head of the cometary globule, B-fields compress into a 
curved morphology by closely following the bright rim of the cometary globule (see R5 model at the top and 
middle panels of Figures 2 and 3 of \citealt{MackeyLim2011MNRAS}) similar to our
present observations towards the BRC (Figure \ref{Bfield_around_BRC}). While 
weak B-field models result in a more significantly aligned B-fields parallel to the pillars, 
strong B-field models ($\sim$160~$\mu$G) produce confined ionized gas to a dense ribbon which standoff from the globule and shields the globule from the
ionizing radiation. In the strong B-field model, B-fields do not exhibit obvious change in their orientation both at the head or tail of the pillar
(see R8 model results at bottom panels of Figures 2 and 3 of \citet{MackeyLim2011MNRAS}).

B-field strength, for the region E associated with the BRC (see Figures \ref{polvecmap_bg} and \ref{vecmap_abcde}), 
is estimated to be 74$\pm$8~$\mu$G. The approximate distance to the BRC from the H{\sc ii} region 
is $\sim$2.2~pc. B-field orientation in a globule prior to the
formation of BRC would be $\sim$163$\degr$ (which corresponds to the dominant 
component of B-fields in the cloud before the onset of star formation, see 
Figure \ref{nirccd_and_ppahk_bffg}(d) and Section \ref{distri_p_t_of_fg_bg}) and is shown with a broken cyan line in Figure \ref{Bfield_around_BRC}. 
The direction of propagating ionizing radiation from the H{\sc ii} region with respect to the BRC 
makes a position angle of 95$\degr$ as shown with a green arrow. 
This implies that initial B-fields around BRC ($\sim 163\degr$) are oriented nearly 
orthogonal to the direction ($\sim 95\degr$) of I-fronts. 
Thus, the estimated B-field strength (74$\pm$8~$\mu$G) and its initial orientation around the BRC 
are similar to those of R5 models of \citet{MackeyLim2011MNRAS} involving perpendicularly oriented B-fields 
with moderate strength (50$\mu$G). 
We believe that this BRC might be formed due to a portion of ionizing radiation that is leaked through 
the low density parts of the molecular cloud, otherwise it would have evaporated if all the radiation from 
H{\sc ii} region is impinged on BRC. 
Therefore, this BRC might experienced the influence of thermal pressure equivalent to that emanated from a single 
O-type star. \citet{MackeyLim2011MNRAS} also used only a single O-type star as the ionizing source in their models. 

According \citet{MackeyLim2013}, at 400 kyr, the models with moderate B-fields ($\sim50~\mu$G), 
vertical motions along the B-fields are initiated (see left panel of Figure 1 of \citealt{MackeyLim2013}). 
Due to this, gas flows in response to photo-evaporation flows along the B-fields result in the formation 
of dense ridge filament elongated parallel to the B-fields.
At 500 kyr a clear cometary globule is formed in which B-fields are aligned with their bright rim similar to our results.
Thus, our results are in accordance with RMHD models of \citet{MackeyLim2011MNRAS,MackeyLim2013} with moderate B-field strength (initially 
oriented perpendicular to the direction of the radiation propagation). Observational evidences similar to our present results includes
compressed B-fields along but behind the bright rim SFO\,74 \citep{Kusuneetal2015}. 

\section{\it Multi-Gaussian fitting on $T$ vs. $V_{LSR}$ spectra}\label{appendix_c}
Initial guess values were provided by visual inspection of observed spectra.
The number of Gaussian components are determined based on the number of peaks appeared in each spectrum.
Since the `gatorplot.pro' does not provide the errors in the fitted multi-Gaussians, we further used an 
output of `gatorplot.pro' as an input to `mpfitpeak.pro' of Markwardt IDL suite of functions \citep{Markwardt2009}
to ascertain the uncertainties to the fitted measurements. 
Instead of varying all three parameters of one Gaussian (peak:T$_{peak}$, center: $V_{LSR}$ and width: $\sigma_{V_{LSR}}$)
we fixed two of them and varied one (similarly, in case of three Gaussians, out of 9 parameters three were varied and 
six were kept constant). Likewise
we constrained all the multi-Gaussian components of each spectrum along with their fitted errors using 
reduced chi-square minimization method.
Equal weights are given to all points of an observed spectrum while fitting was performed. 

\section{\it Column density map using {\it Herschel} data}\label{appendix_d}

Column density map has been constructed using the following 
relation \citep{Kauffmannetal2008}\footnote{see also the document by Jens Kauffmann for more 
details: \url{http://youngstars.nbi.dk/jeskj/AstroSpectra/kauffmann05.pdf}}
\begin{align}
\begin{split}
S^{beam}_{\nu}~=~\left(\frac{N_{H_{2}}}{2.02\times10^{20} {\rmn cm}^{-2}}\right)~\times~\left(\frac{K_{\nu}}{0.01 {\rmn{cm}^{2}} g^{-1}}\right) \\
\times~\left(\frac{\theta_{HPBW}}{10~arcsec} \right)^{2}~\times~\left(\frac{\lambda}{\rmn mm}\right)^{-3} \\
\times~\left(\exp\left({1.439\left(\frac{\lambda}{\rmn mm}\right)^{-1}~\left(\frac{T}{10K}\right)^{-1}}\right)-1\right)^{-1},  
\end{split}
\end{align}\label{sed_eqn}
where $S^{\rmn beam}_{\nu}$ is the flux per beam (mJy~beam$^{-1}$), $N_{{H}_{2}}$ is the column density and $K_{\nu}$ 
(=$0.1\left(\frac{\nu(GHz)}{1000}\right)^{\beta}$; $\nu$ is frequency and 
$\beta$ is dust opacity exponent, and is considered to be 2; \citet{Arzoumanianetal2011,Andersonetal2012}) 
is the dust opacity or specific absorption constant. 
$\theta_{HPBW}$ (arcsec) is beam width (or half-power band width in arcsec) of PACS 160$\mu$m (11$\farcs$4) 
and SPIRE 250 $\mu$m (18$\farcs$1), 350 $\mu$m (24$\farcs$9) and 500 $\mu$m (36$\farcs$4) images. 
$\lambda$ (mm) is wavelength and $T (K)$ is the dust temperature. Pixel units of PACS 160$\mu$m image is Jy/pixel and those of 
SPIRE (250$\mu$m, 350$\mu$m and 500$\mu$m) are MJy/sr. Final pixel units were converted into mJy/beam using the 
following relations\footnote{see page nos 97-100 of dissertation by Dimitrios Stamadianos: 
\url{https://www.escholar.manchester.ac.uk/uk-ac-man-scw:96495}}: 
\begin{eqnarray}
M\left(\frac{\rmn Jy}{\rmn sr}\right)=(10^{-6})\times\left(\frac{\rmn Jy}{\rmn pixel}\right)\times\left(\frac{1}{A_{\rmn pixel}}\right) \\
m\left(\frac{\rmn Jy}{\rmn beam}\right)=(2.665\times10^{-9})~\left(\frac{\rmn Jy}{\rmn pixel}\right)~\times\left(\frac{\theta^2_{HPBW}}{A_{\rmn pixel}}\right),
\end{eqnarray} 
where A$_{\rmn pixel}$ ({\it Sr}) is the pixel area. 
All the images were smoothed to match the resolution of 36$\farcs$4 corresponds to the SPIRE 500$\mu$m image. 
Equation \ref{sed_eqn}1 is fitted to the pixel-wise flux values of four images using `mpfit.pro’ of Markwardt IDL suite of functions \citep{Markwardt2009}. 
Equal weight is given to four input fluxes while fitting is performed. 

\label{lastpage}

\begin{tiny}
\begin{landscape}
\begin{center}
\begin{longtable}{lllllllllllll}
\caption[]{The $JHK_{s}$-band polarimetric and photometric measurements of 1074 stars.}\label{polphotdata1} \\ \hline
\textbf{Star ID} & \textbf{R.A($\degr$)} & \textbf{Dec ($\degr$)} & \textbf{$P(J)$} &  \textbf{$P(H)$}  &  \textbf{$P(K_{s})$} & \textbf{$\theta(J)$}  & \textbf{$\theta(H)$} & \textbf{$\theta(K_{s})$} & \textbf{$J$} &  \textbf{$H$} &  \textbf{$K_{s}$} & Classification   \\ \hline

\endfirsthead
\multicolumn{13}{c}%
{\tablename\ \thetable\ -- \textit{Continued from previous page}} \\

\hline
\textbf{Star ID} & \textbf{R.A($\degr$)} & \textbf{Dec ($\degr$)} & \textbf{$P(J)$} &  \textbf{$P(H)$}  &  \textbf{$P(K_{s})$} & \textbf{$\theta(J)$}  & \textbf{$\theta(H)$} & \textbf{$\theta(K_{s})$} & \textbf{$J$} &  \textbf{$H$} &  \textbf{$K_{s}$} & Classification
\\ \hline

\endhead
\multicolumn{13}{r}%
{\tablename\ \thetable\ -- \textit{Continued on next page}} \\
\endfoot
\endlastfoot
    &   (deg)  & (deg) &  (\%) & (\%) & (\%) & (deg) & (deg)  & (deg)  & (mag) & (mag) & (mag) &  \\
(1)   &  (2)     &   (3)    &   (4)  &  (5)   &   (6)   &  (7)      &  (8)        &   (9)  &  (10) & (11) & (12) & (13)   \\\hline
      1$^\star$   &  167.82417  &  -61.30004  &         ...            &         ...            &   1.54 $\pm$    0.12   &         ...           &         ...           &  156.1 $\pm$    2.2   &  10.926 $\pm$    0.023   &  10.761 $\pm$    0.022   &  10.717 $\pm$    0.021   &      foreground   \\
     2$^\star$   &  167.82453  &  -61.32636  &         ...            &         ...            &   2.39 $\pm$    0.30   &         ...           &         ...           &   56.2 $\pm$    3.6   &  15.745 $\pm$    0.125   &  12.896 $\pm$    0.039   &  11.588 $\pm$    0.052   &      background \\
     3           &  167.82542  &  -61.25781  &         ...            &         ...            &  11.74 $\pm$    6.92   &         ...           &         ...           &  126.7 $\pm$   16.9   &          ...             &          ...             &          ...             &  ...        \\
     4$^\star$   &  167.82617  &  -61.28490  &         ...            &         ...            &   0.74 $\pm$    0.38   &         ...           &         ...           &  125.6 $\pm$   14.7   &  13.251 $\pm$    0.023   &  12.364 $\pm$    0.022   &  12.038 $\pm$    0.024   &                  ...        \\
     5$^\star$   &  167.82662  &  -61.29078  &         ...            &         ...            &   3.81 $\pm$    3.43   &         ...           &         ...           &  135.5 $\pm$   25.8   &  14.972 $\pm$    0.033   &  14.590 $\pm$    0.045   &  14.632 $\pm$    0.101   &                  ...        \\
     6           &  167.82681  &  -61.27721  &         ...            &         ...            &         ...            &         ...           &         ...           &         ...           &  15.709 $\pm$    0.069   &  14.958 $\pm$    0.054   &  14.705 $\pm$    0.062   &     ... \\
     7           &  167.82700  &  -61.27257  &         ...            &         ...            &         ...            &         ...           &         ...           &         ...           &  15.670 $\pm$    0.051   &  14.537 $\pm$    0.032   &  14.029 $\pm$    0.031   &     ... \\
     8           &  167.82750  &  -61.27381  &   2.12 $\pm$    0.07   &   1.16 $\pm$    0.04   &   1.31 $\pm$    0.05   &   39.0 $\pm$    0.9   &   66.3 $\pm$    0.9   &   81.9 $\pm$    1.1   &  11.136 $\pm$    0.006   &  10.148 $\pm$    0.004   &   9.792 $\pm$    0.003   &          background   \\
     9$^\star$   &  167.82789  &  -61.24061  &         ...            &   2.26 $\pm$    0.90   &   2.13 $\pm$    1.23   &         ...           &   81.9 $\pm$   11.4   &   84.8 $\pm$   16.5   &  14.924 $\pm$    0.043   &  13.789 $\pm$    0.029   &  13.331 $\pm$    0.034   &          background   \\
    10           &  167.82793  &  -61.28848  &         ...            &         ...            &   8.42 $\pm$    3.82   &         ...           &         ...           &   38.7 $\pm$   13.0   &  15.629 $\pm$    0.031   &  14.989 $\pm$    0.022   &  14.852 $\pm$    0.035   &      foreground$^\dagger$   \\
    11$^\star$   &  167.82798  &  -61.35974  &         ...            &   9.91 $\pm$    3.27   &   8.22 $\pm$    3.22   &         ...           &   99.8 $\pm$    9.5   &   78.8 $\pm$   11.2   &  17.096                  &  15.273 $\pm$    0.079   &  14.257 $\pm$    0.078   &          background   \\
    12$^\star$   &  167.82850  &  -61.31453  &         ...            &   1.67 $\pm$    1.12   &         ...            &         ...           &   66.7 $\pm$   19.1   &         ...           &  14.636 $\pm$    0.036   &  14.248 $\pm$    0.038   &  14.155 $\pm$    0.063   &                  ...  \\
    13           &  167.82881  &  -61.27652  &         ...            &         ...            &   3.63 $\pm$    1.94   &         ...           &         ...           &   97.7 $\pm$   15.3   &  15.635 $\pm$    0.044   &  14.374 $\pm$    0.021   &  13.893 $\pm$    0.021   &                  ...        \\
    14           &  167.82915  &  -61.30585  &   2.31 $\pm$    1.76   &         ...            &         ...            &   53.0 $\pm$   21.8   &         ...           &         ...           &  14.992 $\pm$    0.020   &  14.057 $\pm$    0.013   &  13.766 $\pm$    0.016   &                  ... \\
    15$^\star$   &  167.82931  &  -61.32771  &         ...            &         ...            &   6.63 $\pm$    2.74   &         ...           &         ...           &   55.5 $\pm$   11.8   &  16.722                  &  15.578 $\pm$    0.164   &  14.359 $\pm$    0.085   &         Class II  \\
    16           &  167.82960  &  -61.27960  &   5.00 $\pm$    2.67   &   2.39 $\pm$    1.98   &         ...            &   30.7 $\pm$   15.3   &   92.2 $\pm$   23.6   &         ...           &  15.393 $\pm$    0.028   &  14.823 $\pm$    0.022   &  14.533 $\pm$    0.029   &                  ...  \\
    17$^\star$   &  167.82972  &  -61.31241  &         ...            &   2.37 $\pm$    1.04   &   3.43 $\pm$    1.48   &         ...           &  100.6 $\pm$   12.6   &   89.2 $\pm$   12.4   &  15.828 $\pm$    0.081   &  14.321 $\pm$    0.055   &  13.752 $\pm$    0.058   &          background   \\
    18           &  167.83039  &  -61.33665  &         ...            &   1.11 $\pm$    0.70   &   3.20 $\pm$    0.75   &         ...           &   79.4 $\pm$   18.1   &  144.0 $\pm$    6.7   &  14.869 $\pm$    0.019   &  13.172 $\pm$    0.009   &  12.342 $\pm$    0.009   &                  ...  \\
    19$^\star$   &  167.83094  &  -61.23723  &         ...            &   6.24 $\pm$    3.39   &         ...            &         ...           &   97.2 $\pm$   15.5   &         ...           &  14.870 $\pm$    0.101   &  14.560 $\pm$    0.114   &  14.444 $\pm$    0.124   &                  ...  \\
    20           &  167.83131  &  -61.33597  &   3.37 $\pm$    0.13   &   4.04 $\pm$    0.04   &   1.46 $\pm$    0.04   &   86.4 $\pm$    1.1   &   85.6 $\pm$    0.3   &   84.6 $\pm$    0.7   &  12.434 $\pm$    0.009   &  10.229 $\pm$    0.003   &   9.275 $\pm$    0.005   &          background   \\
\hline 
\end{longtable}
\end{center}
Notes:\\
Star IDs marked with $\star$ denotes that their photometric data taken from 2MASS \citep{Skrutskie2006} \\
The polarization values given in columns 4--6 are de-biased \\
The uncertainties in the polarization angles given in columns 7--9 are not accounted for overall angular calibration 
uncertainty of 3$\degr$, which is a systematic uncertainty which would affect all the measured polarization angles \\
Column 13: `Foreground' means confirmed foreground stars\\
Column 13: `Background' means confirmed background stars or cluster members\\
Column 13: `Foreground$^\dagger$' means probable foreground stars with $P(H)$~$>$~3\% (see section \ref{secpoleffi})\\
Column 13: `Background$^\dagger$~$^\dagger$' means probable background stars or cluster member with excess polarization (see section \ref{secpoleffi})\\
Column 13: `Background$^\ddagger$' means probable background star or cluster member with depolarization (see section \ref{secpoleffi})\\
The entries those do not have 2MASS photometric uncertainties are left blank\\
A portion of the table is given here, entire table will be available online \\
\end{landscape}
\end{tiny}

\begin{landscape}
\begin{table*}[!ht]
\begin{scriptsize}
	\caption{Central coordinates, widths, number of stars, mean ($\mu_{\theta(H)}$) 
	and dispersion ($\sigma_{\theta(H)}$) in polarization angles (using Gaussian fitting), 
	velocity dispersion ($\sigma_{V_{LSR}}$) using $^{13}$CO(1~--~0) data, volume density ($n(H_{2})$) 
	using {\it Herschel} data and Plummer-like profile fitting on column density maps, and magnetic field strength estimated using Chandrasekhar-Fermi 
	method for the regions A, B, C, D and E.}\label{mfstrength_cf_hilde}
\begin{tabular}{cccccccccccc}\hline \hline
Region &  RA(J2000) &  Dec(J2000) & width in RA & width in Dec & No. of stars & $\mu_{\theta(H)}$  & $\sigma_{\theta(H)}$   &  $\sigma_{V_{LSR}}$ & $n(H_{2})$  & B (CF)$^\dagger$ & B (CF)$^\ddagger$ \\
       &   (deg)     &   (deg)      & (arcmin)    & (arcmin)     &              & (deg)                         &  (deg)    &  (\kms)     &  ($\times$10$^{3}$~cm$^{-3}$) & ($\mu$G) & ($\mu$G) \\
(1)&(2)&(3)&(4)&(5)&(6)&(7)&(8)&(9)&(10)&(11)&(12) \\
\hline
A & 168.0640 & -61.2732 & 1.35 & 1.87 &  7  &  26$\pm$6 & 10 $\pm$ 7 &  0.8$\pm$0.1  & 5.15  $\pm$ 0.68 & 123$\pm$89  & 123$\pm$21 \\
B & 168.0088 & -61.2763 & 1.76 & 2.24 & 24  & 164$\pm$3 & 16 $\pm$ 3 &  1.1$\pm$0.1  & 3.51  $\pm$ 0.41 &  89$\pm$21  &  89$\pm$13 \\ 
C & 167.9485 & -61.2594 & 1.65 & 2.54 & 20  & 153$\pm$2 & 11 $\pm$ 3 &  1.0$\pm$0.2  & 1.01  $\pm$ 0.09 &  63$\pm$23  &  63$\pm$15 \\ 
D & 168.0562 & -61.3415 & 2.99 & 3.48 & 27  & 132$\pm$2 & 14 $\pm$ 3 &  1.4$\pm$0.3  & 2.52  $\pm$ 0.21 & 108$\pm$36  & 108$\pm$27 \\ 
E & 167.8658 & -61.3180 & 1.53 & 1.53 &  6  &  74$\pm$6 & 15 $\pm$ 10 & 0.9$\pm$0.1  & 3.33  $\pm$ 0.52 &  74$\pm$50  &  74$\pm$8 \\
\hline \hline
\end{tabular} \\
Note:
$\dagger$ Errors in B-fields strengths were estimated by propagating uncertainties in $n(H_{2})$, $\sigma_{\theta(H)}$, and $\sigma_{V_{LSR}}$. \\
$\ddagger$ Errors in B-fields strengths were estimated by propagating uncertainties only in $n(H_{2})$ and $\sigma_{V_{LSR}}$. \\
\end{scriptsize}
\end{table*}
\end{landscape}

\begin{table*}[!ht]
\centering
\caption{Multiple Gaussian fitted components of $^{13}$CO brightness temperature ($T$) versus $V_{LSR}$ spectra of
`A', `B', `C', `D', and `E' regions.}\label{gausscomp_abcde}
\begin{tabular}{cccc}\hline \hline
No & $T_{\rmn peak}$  &  $V_{LSR}$  & $\sigma$  \\
   &  (K)                     &  (\kms)                  &  (\kms)  \\
\hline
\multicolumn{4}{c}{Region A}\\
\hline
1$^\dagger$ &    3.38 $\pm$  0.42 &    -26.81 $\pm$ 0.13 &   0.78 $\pm$ 0.12 \\
2 &    4.48 $\pm$  0.28 &    -24.42 $\pm$ 0.16 &   1.60 $\pm$ 0.13 \\
3 &    0.71 $\pm$  0.29 &    -28.75 $\pm$ 0.88 &   1.45 $\pm$ 0.74 \\
4 &    0.76 $\pm$  0.25 &    -20.94 $\pm$ 0.90 &   1.81 $\pm$ 0.69 \\
\hline
\multicolumn{4}{c}{Region B}\\
\hline
1$^\dagger$ &    3.34 $\pm$  0.33 &    -26.46 $\pm$ 0.16 &   1.10 $\pm$ 0.14 \\
2 &    1.43 $\pm$  0.39 &    -19.93 $\pm$ 0.31 &   0.70 $\pm$ 0.29 \\
3 &    1.27 $\pm$  0.28 &    -23.13 $\pm$ 0.48 &   1.28 $\pm$ 0.45 \\
4 &    0.39 $\pm$  0.27 &    -28.95 $\pm$ 1.55 &   1.51 $\pm$ 1.24 \\
5 &    0.74 $\pm$  0.44 &    -18.53 $\pm$ 0.56 &   0.53 $\pm$ 0.46 \\
6 &    0.44 $\pm$  0.45 &    -17.09 $\pm$ 0.81 &   0.49 $\pm$ 0.66 \\
\hline
\multicolumn{4}{c}{Region C}\\
\hline
1$^\dagger$ &    1.80 $\pm$  0.46 &    -26.35 $\pm$ 0.29 &   0.99 $\pm$ 0.24 \\
2 &    0.38 $\pm$  0.28 &    -27.22 $\pm$ 2.21 &   3.27 $\pm$ 2.05 \\
3 &    0.58 $\pm$  0.31 &    -23.18 $\pm$ 1.03 &   1.35 $\pm$ 1.10 \\
4 &    0.69 $\pm$  0.33 &    -19.79 $\pm$ 0.66 &   0.91 $\pm$ 0.66 \\
5 &    0.39 $\pm$  0.45 &    -16.99 $\pm$ 0.78 &   0.48 $\pm$ 0.68 \\
\hline
\multicolumn{4}{c}{Region D}\\
\hline
1$^\dagger$ &    3.16 $\pm$  1.31 &    -25.96 $\pm$ 0.49 &   1.37 $\pm$ 0.34 \\
2 &    1.15 $\pm$  1.19 &    -26.27 $\pm$ 1.45 &   1.72 $\pm$ 1.07 \\
3 &    1.54 $\pm$  0.30 &    -21.24 $\pm$ 0.44 &   1.57 $\pm$ 0.39 \\
4 &    1.15 $\pm$  0.42 &    -23.21 $\pm$ 0.38 &   0.77 $\pm$ 0.35 \\
5 &    0.72 $\pm$  0.29 &    -18.20 $\pm$ 0.80 &   1.34 $\pm$ 0.64 \\
\hline
\multicolumn{4}{c}{Region E}\\
\hline
1$^\dagger$ &    6.05 $\pm$  0.38 &    -23.04 $\pm$ 0.08 &   0.87 $\pm$ 0.07 \\
2 &    3.98 $\pm$  0.30 &    -25.23 $\pm$ 0.14 &   1.19 $\pm$ 0.12 \\
3 &    0.68 $\pm$  0.24 &    -20.34 $\pm$ 1.08 &   2.05 $\pm$ 0.96 \\
\hline \hline
\end{tabular}\\
$^\dagger$ These Gaussian spectra corresponds to the cloud velocity and are
shown with red thick lines in Figure \ref{distri_vel_pa_b}. The gas velocity 
dispersion ($\sigma$) and error corresponding to the cloud component for each region is also 
quoted in the column 9 of Table \ref{mfstrength_cf_hilde}.
\end{table*}

\begin{table*}[!ht]
\centering
\caption{Magnetic and turbulent pressures along with their uncertainties in the regions A, B, C, D and E.}\label{mf_turb_pressure}
\begin{tabular}{cccc}\hline \hline
Region &  $P_{B}$ & $P_{\rmn turb}$ & $P_{B}$/$P_{\rmn turb}$  \\
       &  $10^{-11}$~(dyn~cm$^{-2})$     &  $10^{-11}$~(dyn cm$^{-2})$  &  \\
(1)&(2)&(3)&(4) \\
\hline
A  & 60 $\pm$ 20 &  15 $\pm$   5  &   4 $\pm$   2 \\   
B  & 32 $\pm$  9 &  20 $\pm$   6  &   2 $\pm$   1 \\
C  & 16 $\pm$  8 &   5 $\pm$   2  &   3 $\pm$   2 \\
D  & 46 $\pm$ 23 &  22 $\pm$  11  &   2 $\pm$   1 \\
E  & 22 $\pm$  5 &  11 $\pm$   3  &   2 $\pm$   1 \\
	\hline \hline
\end{tabular} \\
Note: In this table, B-field pressure values and their uncertainties are derived using B-field strength and corresponding uncertainties given in column 12 of
Table \ref{mfstrength_cf_hilde}. 
\end{table*}

\end{document}